\documentclass[fleqn,usenatbib]{mnras}
\pdfoutput=1

\usepackage{newtxtext,newtxmath}

\usepackage[T1]{fontenc}

\DeclareRobustCommand{\VAN}[3]{#2}
\let\VANthebibliography\thebibliography
\def\thebibliography{\DeclareRobustCommand{\VAN}[3]{##3}\VANthebibliography}

\usepackage{graphicx}	
\usepackage{amsmath}	
\usepackage{amssymb}	

\title[Dark matter and the CGM]{The imprint of dark subhaloes on the circumgalactic medium}

\author[I. G. McCarthy \& A. S. Font]{
Ian G. McCarthy,$^{1}$\thanks{E-mail: i.g.mccarthy@ljmu.ac.uk}
Andreea S. Font$^{1}$
\\
$^{1}$Astrophysics Research Institute, Liverpool John Moores University, 146 Brownlow Hill, Liverpool, L3 5RF, UK
}

\date{Accepted XXX. Received YYY; in original form ZZZ}

\pubyear{2020}

\begin{document}
\label{firstpage}
\pagerange{\pageref{firstpage}--\pageref{lastpage}}
\maketitle

\begin{abstract}
The standard model of cosmology, the $\Lambda$CDM model, robustly predicts the existence of a multitude of dark matter `subhaloes' around galaxies like the Milky Way.  A wide variety of observations have been proposed to look for the gravitational effects such subhaloes would induce in observable matter.  Most of these approaches pertain to the stellar or cool gaseous phases of matter.  Here we propose a new approach, which is to search for the perturbations that such dark subhaloes would source in the warm/hot circumgalactic medium (CGM) around normal galaxies.  With a combination of analytic theory, carefully-controlled high-resolution idealised simulations, and full cosmological hydrodynamical simulations (the \texttt{ARTEMIS} simulations), we calculate the expected signal and how it depends on important physical parameters (subhalo mass, CGM temperature, and relative velocity).  We find that dark subhaloes enhance both the local CGM temperature and density and, therefore, also the pressure.  For the pressure and density, the fluctuations can vary in magnitude from tens of percent (for subhaloes with $M_{\rm sub}=10^{10}$ M$_\odot$) to a few percent (for subhaloes with $M_{\rm sub}=10^{8}$ M$_\odot$), although this depends strongly on the CGM temperature.  The subhaloes also induce fluctuations in the velocity field ranging in magnitude from a few km/s up to 25 km/s.  We propose that X-ray, Sunyaev-Zel'dovich effect, radio dispersion measure, and quasar absorption line observations can be used to measure these fluctuations and place constraints on the abundance and distribution of dark subhaloes, thereby placing constraints on the nature of dark matter.
\end{abstract}




\section{Introduction}

The standard model of cosmology, the $\Lambda$CDM model, predicts that every galaxy is surrounded by a dark matter halo within which numerous smaller `subhaloes' exist and orbit about the main system (e.g., \citealt{moore1999,klypin1999,springel2008}).  Conclusively detecting (or refuting) the presence of this large population of dark subhaloes is one of the main quests of modern cosmology, as the abundance and properties of such systems are strongly linked to the physical nature of dark matter (see \citealt{bullock2017} for a recent review).  

A variety of methods have been proposed and are being used to observationally constrain the abundance and properties of the subhaloes, including measurements of the Lyman-alpha forest power spectrum (e.g., \citealt{viel2013}), flux ratio measurements of gravitationally-lensed quasars (e.g., \citealt{mao1998}), perturbations of strong lensing arcs (e.g., \citealt{koopmans2005}), the prevalence of wide-separation stellar binary systems \citep{penarrubia2010}, dynamical heating of thin stellar streams (e.g., \citealt{erkal2015,bovy2017}), among others.  Common to most of these approaches is the distortion (either apparent or physical) of the stellar or cool/neutral phase of matter, which is typically observed in the optical or near-infrared bands.

The increasing fidelity and diversity of measurements of the dominant (by mass) component of baryons, namely the warm/hot phase ($\sim10^{5-8}$ K), raises the interesting question of whether it can be utilised as a probe of dark subhaloes and, more generally, the nature of dark matter.  Specifically, the presence of dark subhaloes orbiting through the circumgalactic medium (CGM) (i.e., the warm/hot phase around galaxies) will exert a gravitational force that will compress the CGM in the vicinity of the subhaloes.  Thus, subhaloes are expected to source fluctuations in the spatial and kinematical properties of the CGM, which may be accessed observationally through common techniques for measuring the CGM, including via diffuse X-ray emission, the thermal and kinetic Sunyaev-Zel'dovich effects, radio dispersion measurements, and via absorption/emission lines in, e.g., quasar spectra.

Here we employ a combination of analytic theory, high-resolution idealised hydrodynamical simulations, and full cosmological hydrodynamical simulations to calculate, for the first time, the expected signal and discuss how it may be maximised.  We are aware of only two previous studies that focused on the local compressive effects of subhaloes on hot diffuse gas, which are those of \citet{churazov2012} and \citet{andrade_santos2013}.  These authors examined how very large infalling subhaloes that host galaxy groups ($M\sim10^{13}$ M$_\odot$) affect the X-ray morphology of the Coma cluster ($M\sim10^{15}$ M$_\odot$).  Here we focus on a very different regime: low-mass dark subhaloes ($10^{7-10}$ M$_\odot$), which, as we will show, will be most easily identified around host galaxies similar to the Milky Way (i.e., that have a significant cool, $\sim10^{5-6}$ K, CGM phase).


This paper is structured as follows.  In Section \ref{sec:analytic} we develop a simple analytic approach for calculating the expected signal for a subhalo that is at rest with respect to the CGM.  In Section \ref{sec:swift} we describe our idealised simulation setup.  In Section \ref{sec:results} we present the main results based on the idealised simulations, demonstrating how the velocity and spatial distribution of the CGM is altered by subhaloes as we systematically vary the important physical parameters of the setup (e.g., subhalo mass, CGM temperature, relative velocity).  In Section \ref{sec:artemis} we analyse the high-resolution \texttt{ARTEMIS} suite of cosmological hydrodynamical zoom-in simulations of Milky Way-mass galaxies, demonstrating the signal is present in realistic, self-consistent simulations.  Finally, in Section \ref{sec:discuss} we summarise and discuss our findings.

\section{Analytic expectations}
\label{sec:analytic}

Before running hydrodynamical simulations of dark subhaloes moving through the CGM, we first try to develop some physical intuition through analytic means.  We consider here a spherical dark matter (sub)halo embedded within the hot CGM, that is at rest with respect to that medium.  While subhaloes are very unlikely to be at rest with respect to the CGM (except perhaps at the apocentres of their orbits), this considerably simplifies the calculation and should still provide the correct qualitative behaviour, in terms of the expected trends for the enhancement in temperature and density due to compression of the gas, so long as the motion of the subhalo with respect to the CGM is subsonic or mildly transonic.

We assume throughout that the CGM is an ideal gas and obeys the adiabatic equation of state, $P = K \rho^{5/3}$, where $P$ and $\rho$ are the pressure and density, respectively, and $K$ is the adiabat (or `entropy').

Initially, in the absence of a dark subhalo perturber, we assume a uniform medium with ambient thermodynamic quantities: $P_{\rm cgm} = P_0$, $\rho_{\rm cgm} = \rho_0$, $K_{\rm cgm} = K_0$, and $T_{\rm cgm} = T_0$.  We now introduce a dark subhalo which, for simplicitly, we assume to have an isothermal mass distribution:
\begin{equation}
\frac{M(<r)}{M_{200}} = \frac{r}{r_{200}} \ \ \ ({\rm for}~r \le r_{200}),\\
\label{eqn:m_r}
\end{equation}
\noindent where $M_{200}$ is the total mass of the subhalo within the radius $r_{200}$ that encloses a mean matter density of 200 times the critical density at $z=0$.  The profile is truncated at $r_{200}$.  

We define the characteristic `virial' temperature (which we use below) associated with this dark subhalo as:
\begin{equation}
\frac{k_B T_{200}}{\mu m_H} = \frac{G M_{200}}{2 r_{200}} \ \ \ ,
\label{eqn:tvir}
\end{equation}
\noindent where $k_B$ is Boltzmann's constant, $G$ is Newton's constant, $\mu$ is the mean molecular weight, and $m_H$ is the mass of a hydrogen atom.

Locally, within the gravitational sphere of influence of the subhalo, the CGM will strive to obtain a condition of hydrostatic equilibrium, such that:
\begin{equation}
\frac{dP_{\rm cgm}(r)}{dr} = -\frac{G M_{\rm grav}(<r) \rho_{\rm cgm}(r)}{r^2} \ \ \ ,   
\label{eqn:hse}
\end{equation}
\noindent where $r$ is the distance from the subhalo centre and $M_{\rm grav}$ is the subhalo's total enclosed mass profile, for which we ignore the generally negligible additive contribution of the CGM.

One may question the validity of the hydrostatic approximation, as typically the velocity of the subhalo is of order the circular velocity of the host, which is generally larger than the escape speed of the satellite.  Consequently, the CGM near the subhalo will generally not be bounded to the subhalo.  However, so long as motion of the satellite remains relatively slow (transonic or subsonic), the hydrostatic approximation will not be greatly in error, as we confirm with the idealised simulations presented below.  Note that eqn.~\ref{eqn:hse} is derived from the fluid momentum equation (i.e., conservation of momentum) in the limit where the acceleration of the fluid in the presence of the subhalo is small compared to the sum of the gravitational acceleration and the gas pressure gradient.  As demonstrated by \citet{andrade_santos2013}, this will be case for subsonic and mildly transonic motion (see the appendix of that study), which is typically the case for satellites orbiting through the CGM.  Specifically, the orbital velocity is typically the circular velocity of the host, $\sqrt{G M_{200} / r_{200}}$, while the CGM sound speed is typically $\sqrt{k_B T_{200} / (\mu m_H)}$, as defined in eqn.~\ref{eqn:tvir}.  The typical Mach number then is $\approx \sqrt{2}$, which is transonic.  Consequently, the acceleration of the fluid will generally be small and a condition of approximate hydrostatic equilibrium will result.

If we assume that the process of gravitationally squeezing of the CGM near the subhalo is an adiabatic one, such that $K(r)=K_0$, one arrives at:
\begin{equation}
K_0 \frac{d\rho_{\rm cgm}^{5/3}}{\rho_{\rm cgm}} = -\frac{G M_{\rm grav}}{r^2} dr \ \ \ ,
\label{eqn:hse_2}
\end{equation}
\noindent where we have replaced the pressure using the adiabatic equation of state and extracted the constant entropy from the differential on the left-hand side.  We have also removed the explicit radial dependence from $\rho_{\rm cgm}$ and $M_{\rm grav}$ for clarity, but note that these quantities are radially dependent.

Replacing $M_{\rm grav}$ in eqn.~\ref{eqn:hse_2} using eqn.~\ref{eqn:m_r} and integrating with respect to radius, we obtain
\begin{equation}
K_0 \int_r^R \rho_{\rm cgm}^{-1/3} d\rho_{\rm cgm} = -\frac{3}{5} \frac{G M_{200}}{r_{200}}\int_r^R \frac{dr}{r} \ \ \ ,
\label{eqn:hse_int}
\end{equation}
\noindent where $R$ is the radius at which the thermodynamic quantities return to their ambient (unperturbed) values [e.g., $\rho_{\rm cgm}(R)=\rho_0$].  Note that if one neglects the gravitational influence of the host halo in which the the subhalo orbits (as we do in this simple calculation), then $R \sim r_{200}$.  In fact, our derivation above assumes that $R \le r_{200}$.  Embedded within a more massive host halo, we expect that $R$ should be of order the tidal radius of the subhalo (typically $\ll r_{200}$), which is approximately the radius where the mean enclosed density of the subhalo is of order that of the local (total) density of the host halo \citep{king1966}.

Evaluating the integrals and after some algebra, we obtain:
\begin{equation}
\rho_{\rm cgm}(r)^{2/3} - \rho_0^{2/3} = \frac{2}{5} \frac{G M_{200}}{r_{200}} \frac{1}{K_0} \ln{\biggl(\frac{R}{r}\biggr)} \ \ \ .
\end{equation}

Finally, dividing through by $\rho_0^{2/3}$ and using eqn.~\ref{eqn:tvir} and the ideal gas law, we derive the radial density profile of the CGM centred on the subhalo:
\begin{equation}
\frac{\rho_{\rm cgm}(r)}{\rho_0} = \biggl[1 + \frac{4}{5} \frac{T_{200}}{T_0} \ln{\biggl(\frac{R}{r}\biggr)}\biggr]^{3/2} \ \ \ ({\rm for}~r \le R).    
\label{eq:rho_prof}
\end{equation}

Generally speaking, we expect $T_{200}$, the subhalo's characteristic virial temperature, to be less than the CGM temperature of the host halo.  Thus, the additive term due to the compression will generally only result in a small enhancement in the density.  Note that the ambient (unperturbed) value of $\rho_{\rm cgm} = \rho_0$ is correctly recovered in the limit of a massless subhalo, such that $M_{200} \propto T_{200}^{3/2} \rightarrow 0$ and the additive term vanishes.  

Adopting a more realistic NFW form would give rise to a more complex radial dependence, such that:
\begin{equation}
\ln{\biggl(\frac{R}{r}\biggr)} \rightarrow \frac{\frac{R}{r}\ln{\biggl(1+\frac{r}{r_s}\biggr)} - \frac{B}{1+B}}{\ln{(1+B)}-\frac{B}{1+B}} \ \ \ ({\rm for}~r \le R),
\end{equation}

\noindent where $B \equiv R/r_s$ and $r_s$ is the NFW scale radius (i.e.,the radius where the logarithmic slope of the density is $-2$).  

As the compression of the gas is adiabatic, the pressure and temperature profiles are related to eqn.~\ref{eq:rho_prof} simply by:
\begin{equation}
    \frac{P_{\rm cgm}(r)}{P_0} = \biggl(\frac{\rho_{\rm cgm}(r)}{\rho_0}\biggr)^{5/3} =
    ~\biggl(\frac{T_{\rm cgm}(r)}{T_0}\biggr)^{5/2}  \ \ \ .
\label{eqn:p_t_prof}    
\end{equation}

These relations hold independent of the assumption of hydrostatic equilibrium or the form of the gravitational potential and assume only that the gas is ideal and that the compression of the CGM is an adiabatic process. 

From eqn.~\ref{eqn:p_t_prof} it is apparent that the pressure is the most strongly affected thermodynamic quantity, followed by the density, and then the temperature.  As a consequence, observables that probe the gas pressure, such as the thermal Sunyaev-Zel'dovich (tSZ) effect, are expected to show a larger change (in relative terms) than observables that sample the density, such as quasar absorption line studies or radio dispersion measurements.  Perhaps most promising, however, is the X-ray surface brightness, which scales with the density squared and also has a weak temperature dependence (depending on the X-ray instrument energy range and the temperature of the CGM).

We note that the derived profiles imply that the fractional change in the thermodynamic quantities depends only on distance from the subhalo and the ratio of the virial temperature of the subhalo to that of the CGM.  So, for example, the local temperature enhancement of the CGM near the dark subhalo is independent of the density of the CGM.  

The above results were derived for the case of a subhalo which is at rest with respect to the CGM.  We have confirmed their accuracy by comparison with an idealised hydrodynamical simulation in which the subhalo is at rest with respect to the CGM (this simulation is discussed below, in Section \ref{sec:results}).  Introducing a velocity with respect to the CGM will affect the details of the compression and alter the spatial morphology of the gas near the subhalo and requires detailed hydrodynamical simulations to accurately calculate the effect, as we do below.  We also expect that introducing the gravitational influence of the host halo (which we neglected above) will impact the resulting morphology of the CGM near the subhalo, as gas that is relatively far from the subhalo will be more strongly influenced by the main halo (or even by other subhaloes).  In Section \ref{sec:artemis} we will examine the compression of the CGM near subhaloes in the \texttt{ARTEMIS} suite of full cosmological hydrodynamical simulations.

\section{Idealised simulations}
\label{sec:swift}

In this section we describe the setup of the idealised simulations of subhaloes moving through the CGM.  We use idealised (i.e., non-cosmological) simulations as it allows us to carefully control different aspects of the setup, allowing us to isolate the physics at play in a straightforward way.  It also means we can push to significantly higher resolution than is typical in most cosmological simulations.

\subsection{The SWIFT code}

To carry out the hydrodynamical simulations, we use the SWIFT code\footnote{\url{http://www.swiftsim.com}} \citep{schaller2018}.  SWIFT is a massively-parallel hydrodynamics and gravity code for astrophysics and cosmology, which employs a hybrid MPI + threads parallelisation scheme. SWIFT uses the Fast Multipole Method to evaluate gravitational forces between nearby particles and uses a standard particle-mesh algorithm to compute long-range forces.  For our default setup (described below), however, we represent the dark subhaloes with static gravitational potentials, for which a variety of forms are already available as options within the code.  For these idealised simulations we neglect the self-gravity of the gas by default.  However, we have verified that there are no significant differences (other than a much longer run time) when self-gravity is included.

To compute hydrodynamic (pressure) forces, SWIFT uses smoothed particle hydrodynamics (SPH), which is implemented in a variety of ways, including the scheme used in the Gadget-2 code (\citealt{springel2002}, so-called `density--entropy' SPH, where the density is the SPH-smoothed quantity and entropy is the independent thermodynamic quantity), a `pressure--entropy' formulation proposed by \citet{hopkins2013}, the `GIZMO' meshless finite volume (MVF) and finite mass (MFM) solvers introduced by \citet{hopkins2015}, and a new `density--energy' formulation developed for the EAGLE-XL project called `SPHENIX' (Borrow et al., in prep).  By default, we adopt the SPHENIX flavour of SPH solver, but we have verified for our simple physical setup that the choice of hydro solver is unimportant (see Appendix \ref{sec:hydro_study}).  For specificity, the SPHENIX scheme includes the time-step limiter of \citet{durier2012}, an implementation of thermal diffusion following \citet{price2012}, and `inviscid' artificial viscosity with a Balsara switch following \citet{cullen2010}.  Finally, for the smoothing kernel we use a higher-order quintic spline whose width is equivalent 48 neighbours in the standard SPH cubic spline. 

Finally, a wide range of subgrid models for galaxy formation (e.g., radiative cooling, star formation, stellar and AGN feedback, etc.) are also available within SWIFT, including the full EAGLE model \citep{schaye2015}.  However, for simplicity and to isolate the effects we wish to study, we run non-radiative idealised simulations only.  We do not expect the exclusion of galaxy formation physics to impact our results or conclusions in any case, since we focus only on the gravitational effects of dark subhaloes (i.e., that lack galaxies) on neighbouring hot gas that constitutes the CGM.  In Section \ref{sec:artemis} we use the ARTEMIS simulations to verify that our conclusions are robust to the inclusion of galaxy formation physics within a full cosmological context.

\subsection{Fiducial setup}

The default setup we employ is as follows.  The simulation volume is chosen to be a static Newtonian (i.e., non-expanding) periodic cuboid with dimensions ($L_x$, $L_y$, $L_z$) = (1500 kpc, 300 kpc, 300 kpc).  We have selected this dimensionality to mimic a `wind tunnel', in which the hot gas will flow along the $x$-axis over top of the dark subhalo.  This setup means we avoid wasting computational effort simulating the medium at large distances perpendicular to the direction of motion, allowing us to adopt very high mass resolution even on modest computational resources.

The box is initially filled with a uniform density and temperature medium of gas particles, populated using a 3-D regular grid.  By default, the medium has a temperature of $T=10^6 K$ and an atom number density of $n=10^{-4}$ cm$^{-3}$, values typical for the CGM of Milky Way-mass ($M_{\rm halo} \sim 10^{12}$ M$_\odot$) host haloes (e.g., \citealt{gupta2012}).  However, we vary both the temperature and density below.  By default, we adopt an SPH particle mass of $1000$ M$_\odot$ which, given the atom number density, yields just over 200 million SPH particles in the simulation volume.  The typical SPH smoothing length is $\approx 1$ kpc.  Note that this resolution is considerably higher than can be achieved in most cosmological simulations of the CGM (a notable exception is the recent study of \citealt{vandevoort2019}, which also achieved 1 kpc resolution).  When varying the density of the medium (e.g., lowering it to $n=10^{-5}$ cm$^{-3}$), we keep the number of SPH particles fixed and simply vary the SPH particle mass.  In practice, though, we find the results do not depend on the adopted density of the CGM (see Appendix \ref{sec:dens_dependence}), as anticipated from the analysis in Section \ref{sec:analytic}).

To mimic the motion of a dark subhalo through the CGM, the dark subhalo static potential is placed at the centre of the cuboid and the gas particles are given a uniform velocity along the $x$-axis.  By default, the velocity is chosen to be $200$ km/s (i.e., of order the circular velocity of the Milky Way), but we vary this quantity as well.  Note that, while we choose default values for the CGM physical parameters that may be typical for a Milky Way-mass host, our setup applies generally to all galaxies which have a CGM component.  By varying the parameter values as we do, we are effectively surveying the CGM over a wide range of host halo masses.

Dark subhaloes are represented by default as static NFW potentials, which are characterised by their total mass, $M_{200}$, and concentration, $c_{200}$.  (Note that the static potentials are truncated beyond $r_{200}$.)  We convert this into an effectively single parameter potential by using the mass--concentration relation of \citet{dutton2014}.  

We note that we have also explored using fully live potentials, where the dark matter subhalo is represented with a large number of collisionless dark matter particles in Jeans equilibrium.  For this, we set up equilibrium haloes following the methods described in \citet{mccarthy2007} and we used identical masses for the gas and dark matter particles in order to minimize two-body heating effects.  We found fully consistent results with our default static potential runs (which are considerably cheaper).  Naively, one might have expected some energy transfer from the dark subhalo into the CGM via dynamical friction for the live halo runs.  However, we found such dynamical heating to be minimal, which we surmise is due to the low subhalo masses we are considering,  that are embedded within a hot medium.  It is plausible that dynamical heating would become more relevant for higher-mass subhaloes and/or a significantly cooler gaseous medium.

By default we select a mass of $M_{200} = 10^{10}$ M$_\odot$ for the subhalo, which is the highest mass we explore and for which the effects will be most obvious.  Note that haloes of this mass are not strictly speaking `dark' and are thought to host low-mass galaxies with typical stellar masses of $M_{\rm star} \sim 10^{5-7}$ M$_\odot$ (e.g., \citealt{sawala2015,hopkins2018}).  Such systems are nevertheless an interesting test case, though, as they are unlikely to retain their own hot gas gas if embedded within the CGM of a much more massive host.  This is a consequence of efficient ram pressure stripping that results from the motion of the subhalo with respect to the CGM of the host (e.g., \citealt{mccarthy2008,phillips2015}).  Indeed, satellite dwarf galaxies of this stellar mass within the Milky Way are essentially devoid of even denser, cooler atomic gas (e.g., \citealt{grcevich2009,spekkens2014}) which would be considerably more difficult to strip than diffuse, hot gas.

Finally, all simulations are run for a duration of 2 Gyr.  We find, however, that a steady-state configuration is rapidly obtained and for the plots presented below, we show the results after 0.5 Gyr of evolution.

\section{Results}
\label{sec:results}

In this section we present the main results of our study.  We first examine the fiducial setup, before systematically varying the main physical parameters of the idealised simulations.

\subsection{Fiducial simulation}

\begin{figure}
  \includegraphics[width=0.499\columnwidth]{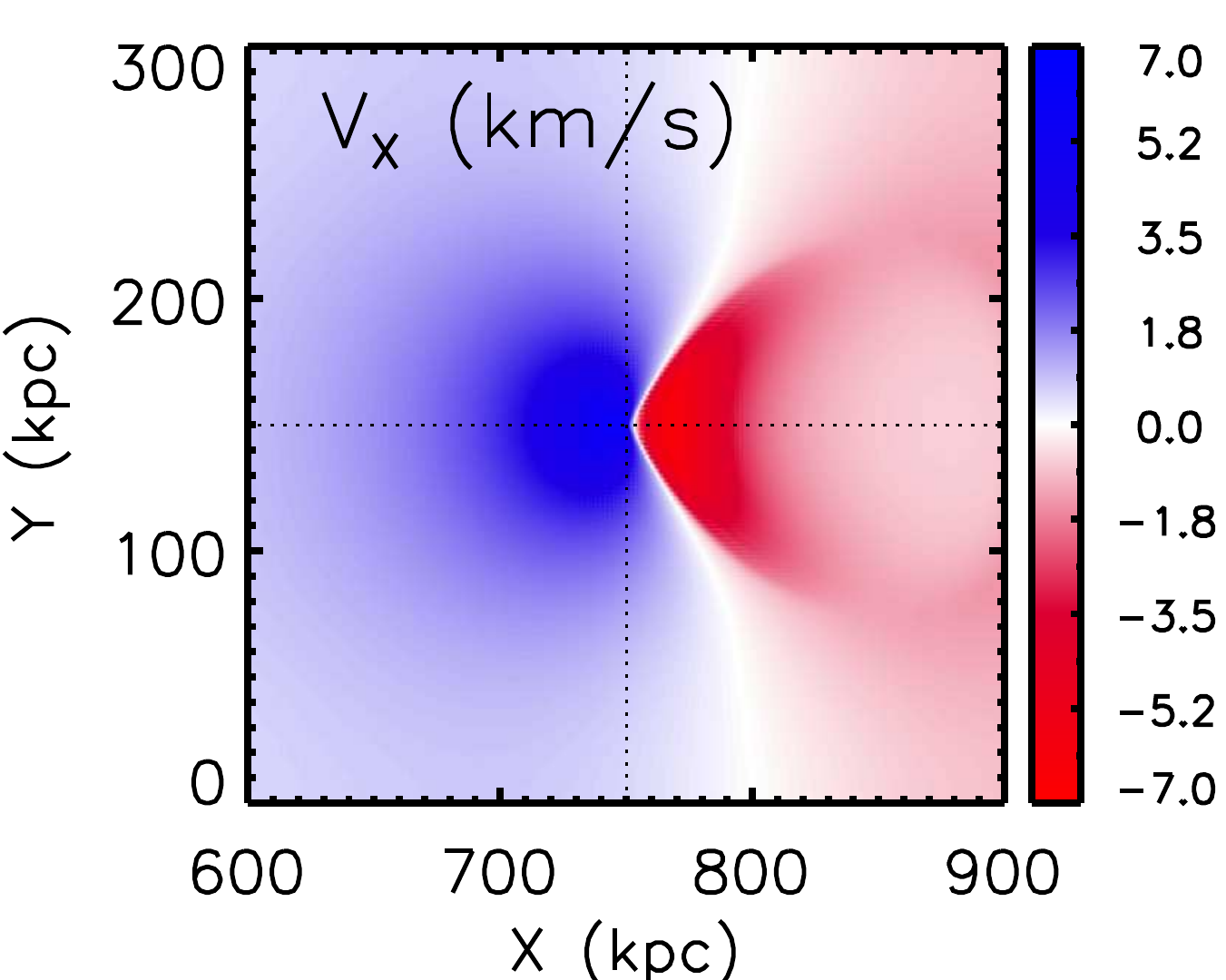}
  \includegraphics[width=0.499\columnwidth]{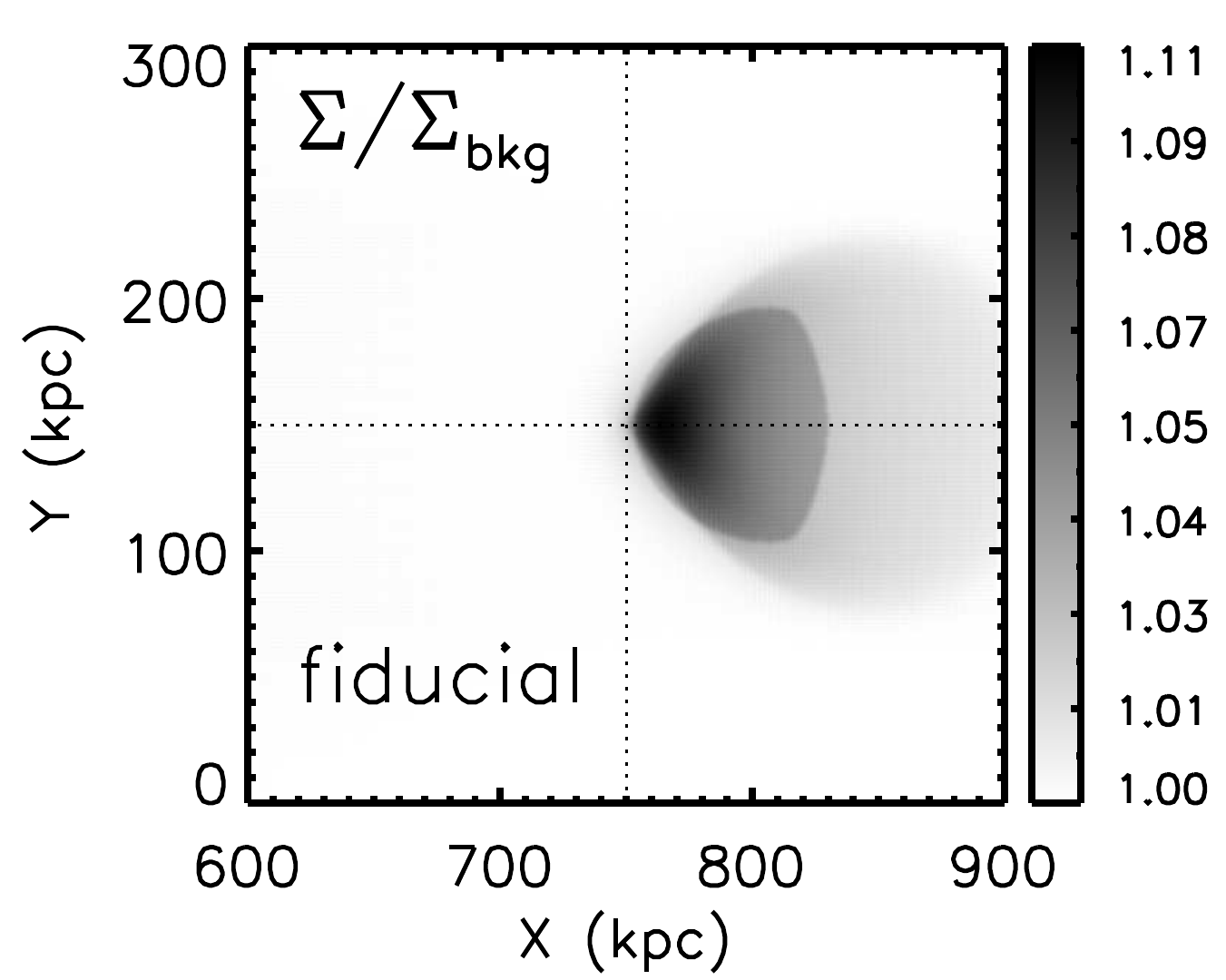}\\
  \includegraphics[width=0.499\columnwidth]{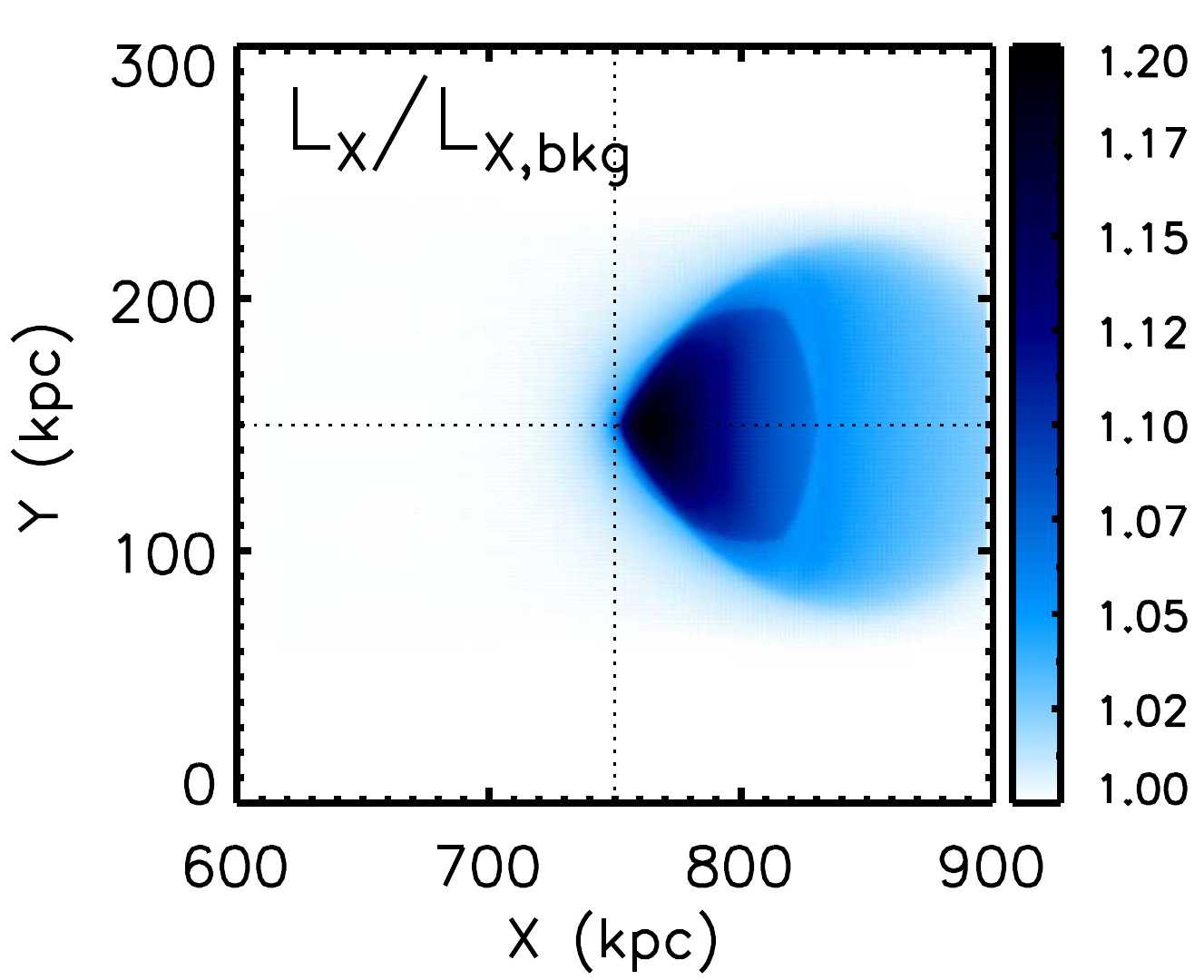} 
  \includegraphics[width=0.499\columnwidth]{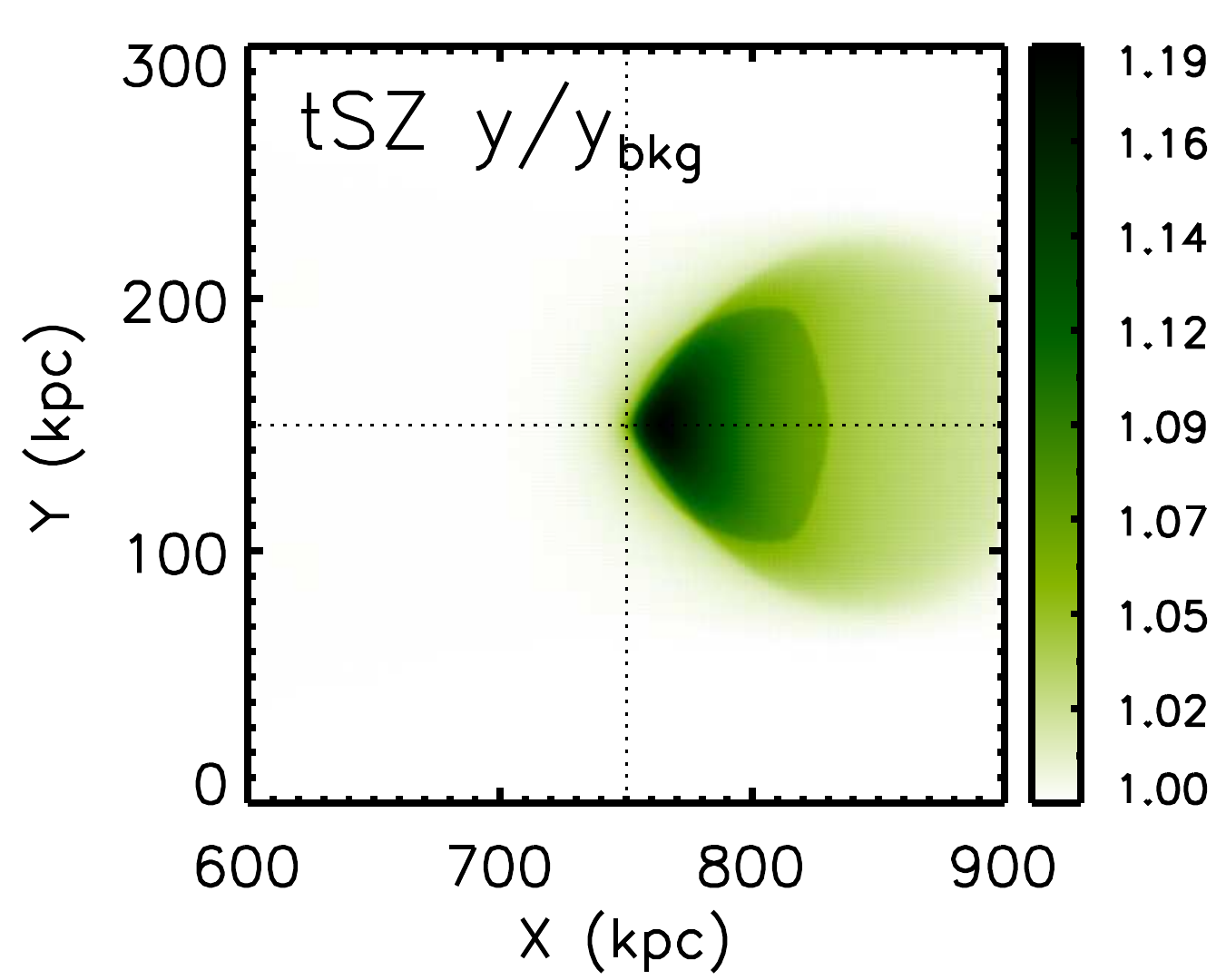}
  \caption{Projected $x$ velocity (top left), surface mass density (top right), X-ray surface brightness (bottom left), and tSZ Compton $y$ parameter maps for the fiducial setup for a side-on view (such that the suhbhalo is moving from right to left).  For all panels aside from the velocity, the quantities have been normalised by the ambient background quantities, therefore showing the relative enhancement.  Where the vertical and horizontal dotted lines cross represents the centre of the subhalo.  
  }
  \label{fig:fiducial}
\end{figure}

In Fig.~\ref{fig:fiducial} we show a variety of projected maps of the CGM, including the density-weighted $x$ component of the velocity (top left), surface mass density (top right), X-ray surface brightness (bottom left) and tSZ effect (bottom right).  In all cases we show an edge-on configuration, such that the subhalo is moving from right to left.  To create projected maps we integrate along the $z$-axis over a depth of 100 kpc (i.e., $|z-z_{\rm subhalo}| < 50$ kpc).  We normalise the surface mass density, X-ray surface brightness, and tSZ maps by the background value, which we calculate using the unperturbed medium from the initial conditions (alternatively, the background could be estimated far from the subhalo).  

To compute the X-ray surface brightness and tSZ effects maps, we follow the methods described in \citet{mccarthy2017} except that here we use the `bolometric' X-ray surface brightness (0.02-50 keV band), as opposed to a soft 0.5-2.0 keV estimate.  As our idealised simulations have no star formation or stellar evolution, we simply assume the metallicity of the gas to be 0.3 $Z_\odot$ when computing the X-ray surface brightness.  Note that, as we are mainly interested in the local enhancement of the X-ray and tSZ effect relative to the unperturbed background, as opposed to absolute surface brightnesses, the exact details of the passband and metallicity are generally unimportant. 

Examining the surface mass density, X-ray surface brightness, and tSZ maps, we find they are just scaled versions of each other (i.e., have identical morphologies), as expected for adiabatic compression and from the analysis in Section \ref{sec:analytic}.  Also consistent with the analytic expectations is that the X-ray surface brightness is the most strongly affected quantity, followed by the tSZ effect, and then the surface mass density.  For this particular setup and projected through a column of 100 kpc, the local enhancement in the X-ray surface brightness and tSZ effect can be up to $\approx20\%$, whereas the surface mass density is enhanced by up to $\approx10\%$.  

When viewed from this orientation (i.e., perpendicular to the direction of the subhalo's motion), the CGM density/temperature/pressure is mostly strongly perturbed slightly downstream from the centre of the subhalo and displays a cone-like (or, in 2-D, a `guitar pick') morphology. 

\begin{figure}
    \includegraphics[width=\columnwidth]{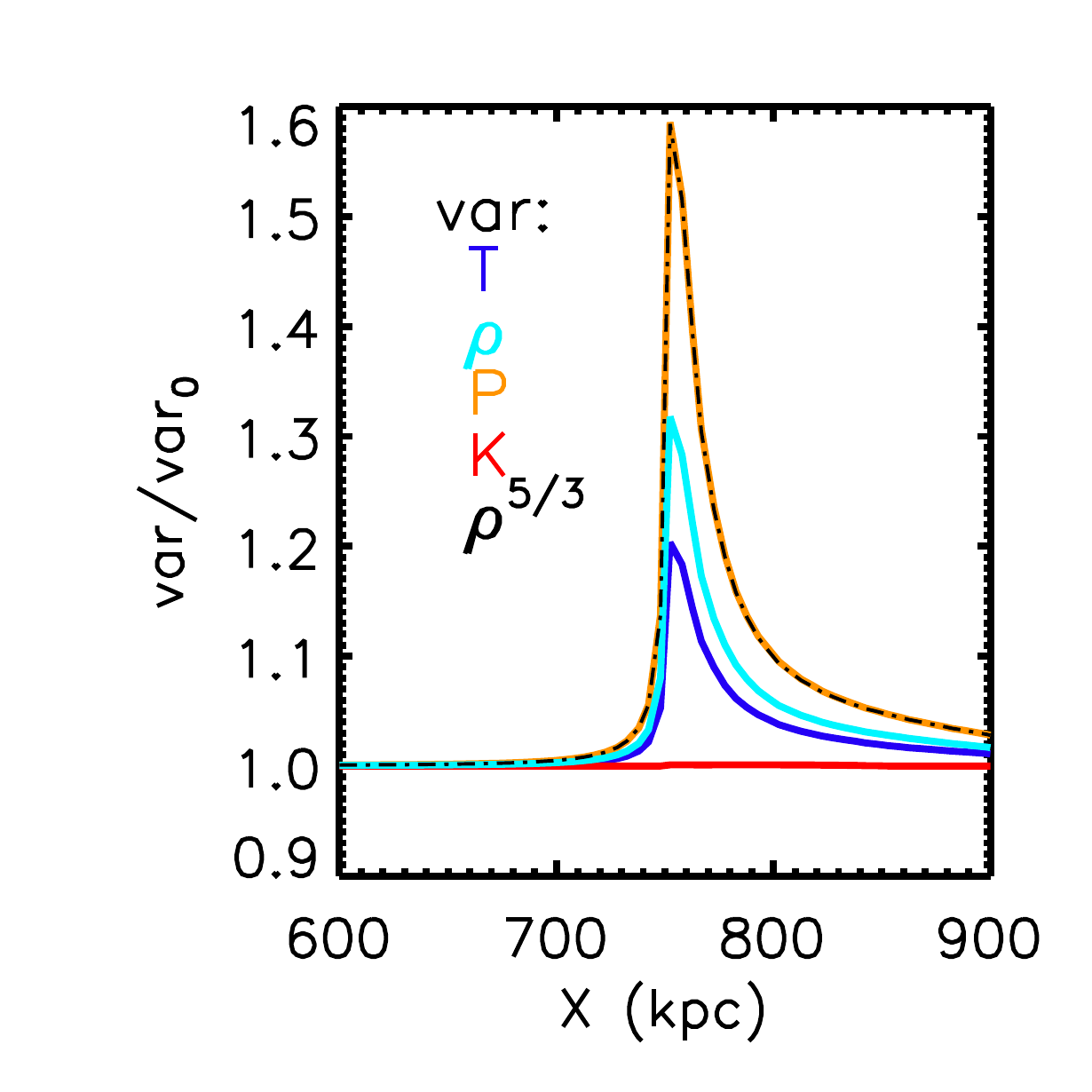}
    \caption{1-D profiles of the temperature (blue), density (cyan), pressure (orange), and entropy (red) along the $x$-axis within a narrow cylinder of width 2.5 kpc (i.e., $r_{yz} < 2.5$ kpc) for the fiducial simulation. The dot-dashed black curve (on top of the orange curve) shows the density to the 5/3 power, confirming the adiabatic nature of the interaction. }
    \label{fig:fiducial_profs}
\end{figure}

\begin{figure}
  \includegraphics[width=0.499\columnwidth]{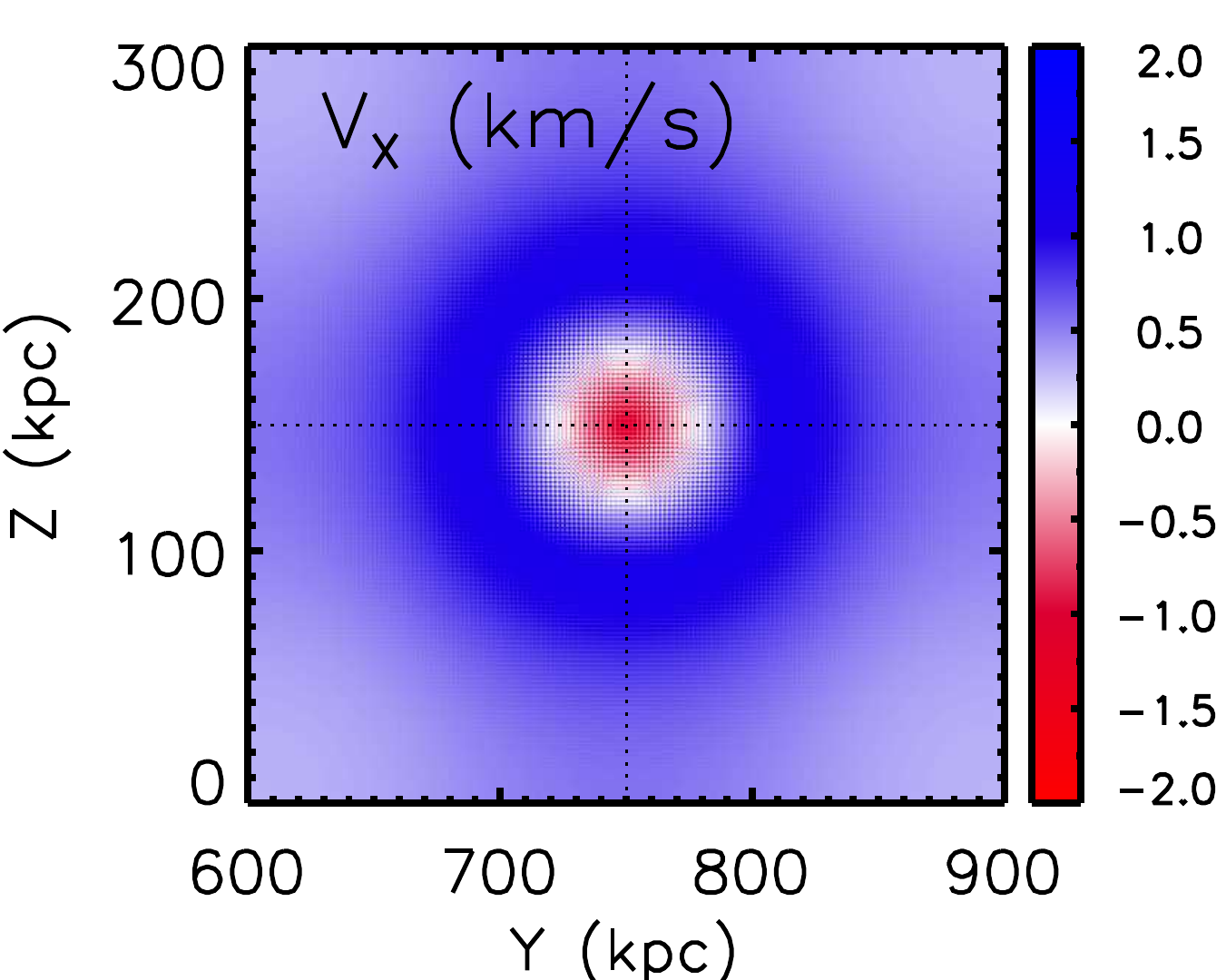}
  \includegraphics[width=0.499\columnwidth]{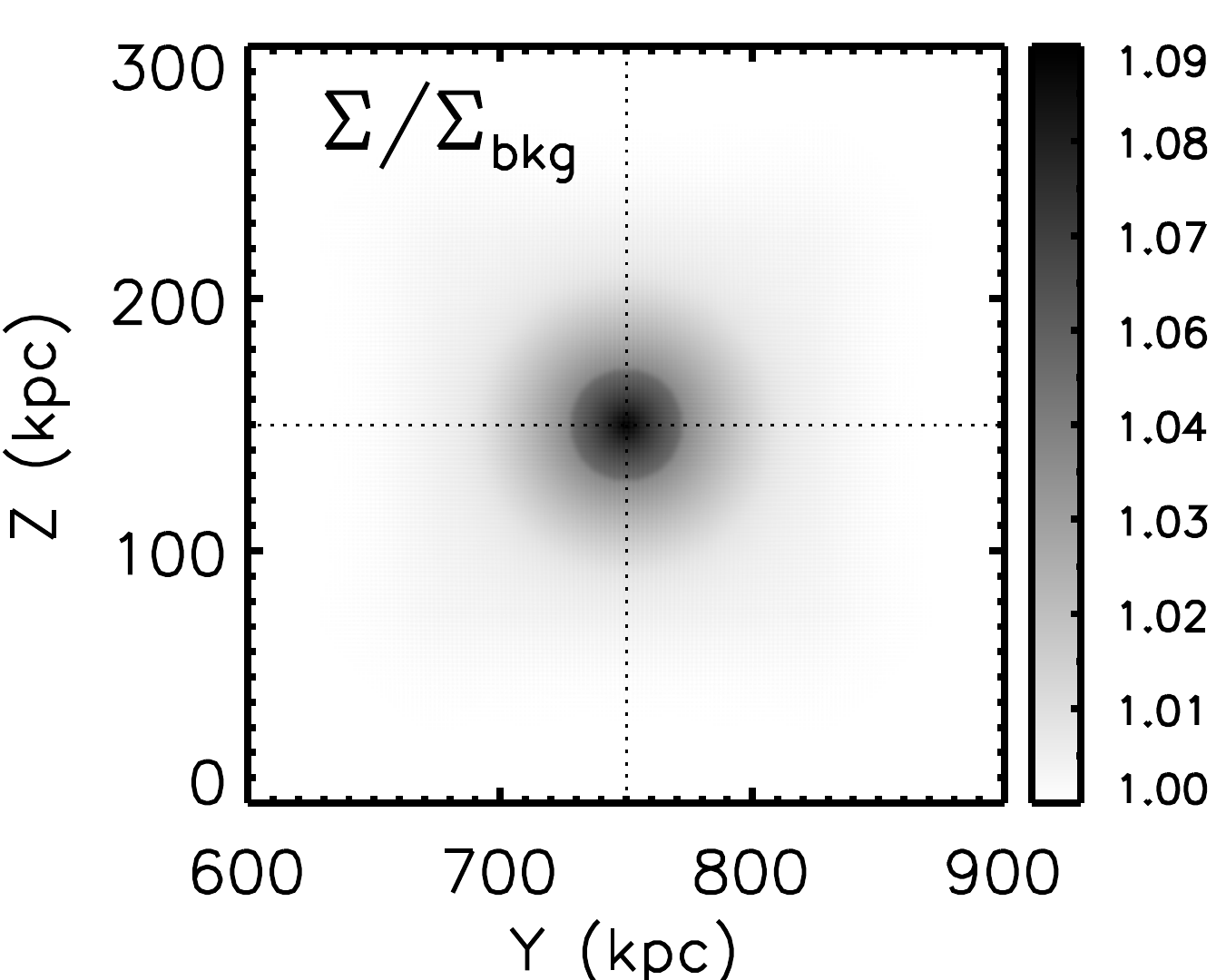}
  \caption{Projected velocity (top left) and surface mass density (top right) maps for the fiducial setup for a face-on view (such that the suhbhalo is moving into the page).    
  }
  \label{fig:fiducial_face_on}
\end{figure}

The velocity map shows understandable features as well.  Gas ahead of the oncoming subhalo (left side) accelerates towards the subhalo (positive $x$ velocity, displayed in blue) and has a smooth, almost circular geometry.  Gas that has passed through the subhalo and is now trailing behind it (right side), is dragged along by the subhalo with a (projected) velocity in this case of up to $\approx -7$ km/s (negative $x$ velocity, displayed in red).  Note that the subhalo is moving with a velocity of $-200$ km/s, so the gas is clearly unbound from the subhalo and will come to rest further downstream.  Note also that, as we are examining here the $x$-component of the velocity, we are examining the (projected) transverse velocity in this case and not the line-of-sight velocity.

In Fig.~\ref{fig:fiducial_profs} we show 1-D profiles along the $x$-axis of the four thermodynamic variables (noting only two are independent).  The profiles were constructed by selecting only particles that lie within a cylinder of width 2.5 kpc down the $x$ axis through the middle of the box.  We then take the median value of particles in bins of width $\Delta x = 5$ kpc.

As expected for this regime of subsonic/transonic motions, the assumption of adiabatic compression is well justified.  Little to no entropy is generated through shocking or mixing.  As an important caveat, however, we note that our simulations do not include radiative cooling.  The compression of the CGM as it passes near the subhalo will temporarily lead to enhanced cooling rates (thus reducing the entropy), scaling as approximately the square of the density.  If the CGM is particularly close to being thermally unstable, i.e., if the ratio of the local cooling time to the free-fall time is $\lesssim10$ (e.g., \citealt{mccourt2012}), the perturbations introduced by the subhaloes could induce run-away cooling, leading to the formation of cold gas (such as `high-velocity clouds') and potentially even star formation.  We neglect this potentiality here, noting that the majority of the CGM is not expected to be finely-perched on this threshold.  If anything, however, such cooling would only emphasise the observability of these fluctuations (as the density increases as a result of the cooling) and is therefore worth future study.

Fig.~\ref{fig:fiducial} shows maps when viewing the subhalo from a direction that is perpendicular to its direction of motion.  In Fig.~\ref{fig:fiducial_face_on}, we show the analogous velocity and surface mass density maps but for a face-on configuration, where the subhalo is moving directly along the line of sight (into the page).  As the X-ray surface brightness and tSZ effect maps look identical to the surface mass density map (but with different amplitudes), we omit them here.

The surface mass density map (right panel) shows a circularly-symmetric distribution with a peak amplitude that is similar to that of the edge-on configuration.  The central region is significantly enhanced compared to the outer regions and demarcates the spatial extent of the strong conical feature evident in Fig.~\ref{fig:fiducial}.

The velocity map (left panel) shows strongly reduced velocities overall, a result of the fact that we are now examining the line of sight velocity for which there is some cancellation; i.e., gas upstream of the subhalo moves with a positive $x$ velocity (out of the page) while gas downstream of the subhalo moves in the opposite direction.  The strong density enhancement downstream (the cone) means that that cancellation is not perfect, such that the the central regions show a net negative velocity while the outer regions show a net positive velocity.  If the subhalo were moving in the opposite direction (out of the page), the map colours would be inverted.

Finally, it is interesting to note that density fluctuations of similar amplitudes ($\sim5$--$20\%$) and on similar spatial scales ($\sim10$ kpc) have been observed in the central regions of the Perseus cluster using deep {\it Chandra} X-ray data (\citealt{zhuravleva2015}, see also \citealt{zhang2009}).  In that particular case the fluctuations are likely the result of turbulence driven by AGN feedback, rather than compression due to dark subhaloes.  Nevertheless, it shows that fluctuations of this amplitude are already potentially observable with current instrumentation.  Future deep observations with {\it Lynx}\footnote{\url{https://wwwastro.msfc.nasa.gov/lynx/}} and {\it Athena}\footnote{\url{https://www.the-athena-x-ray-observatory.eu/}} should therefore have tremendous constraining power.

\begin{figure}
  \includegraphics[width=0.499\columnwidth]{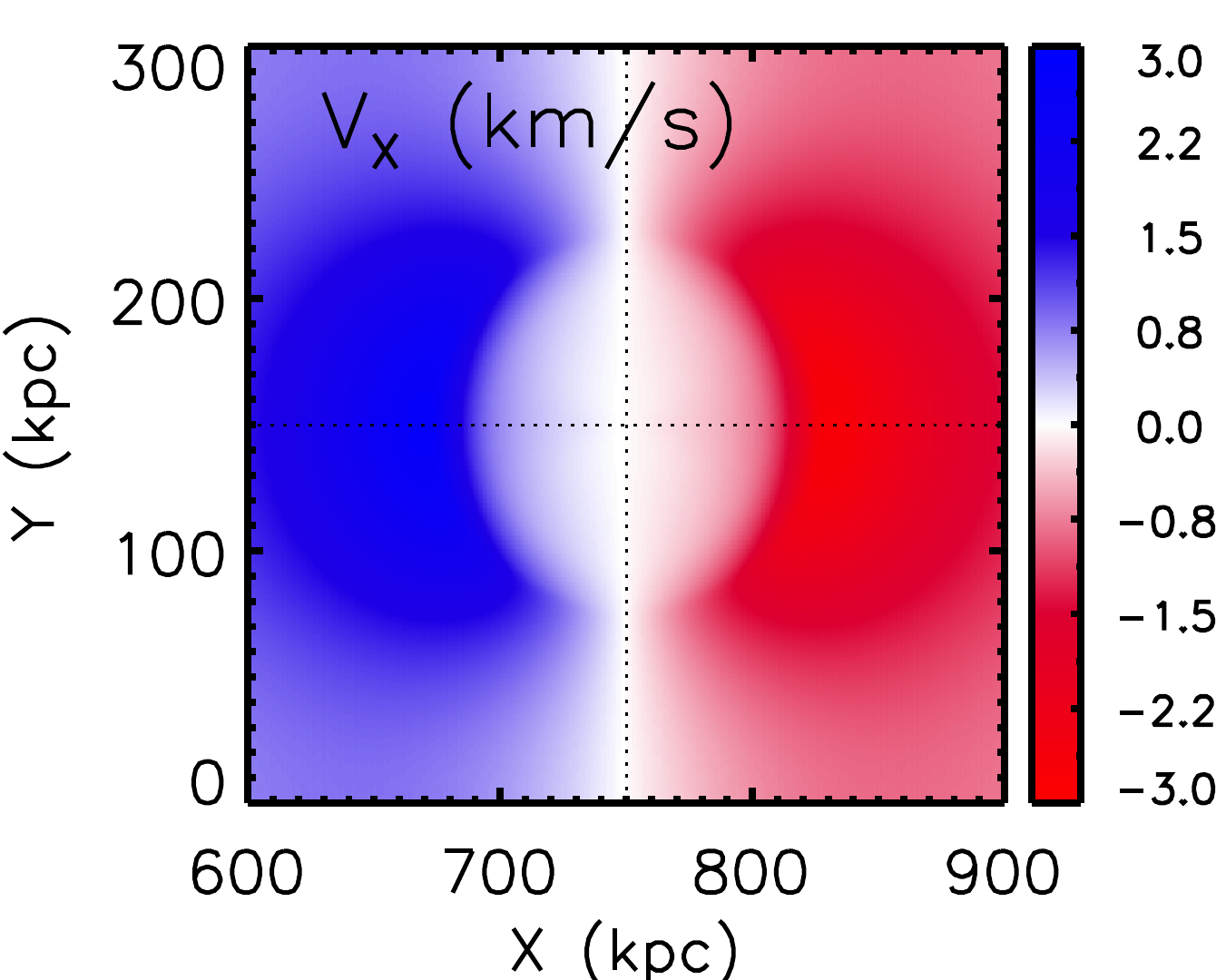}
  \includegraphics[width=0.499\columnwidth]{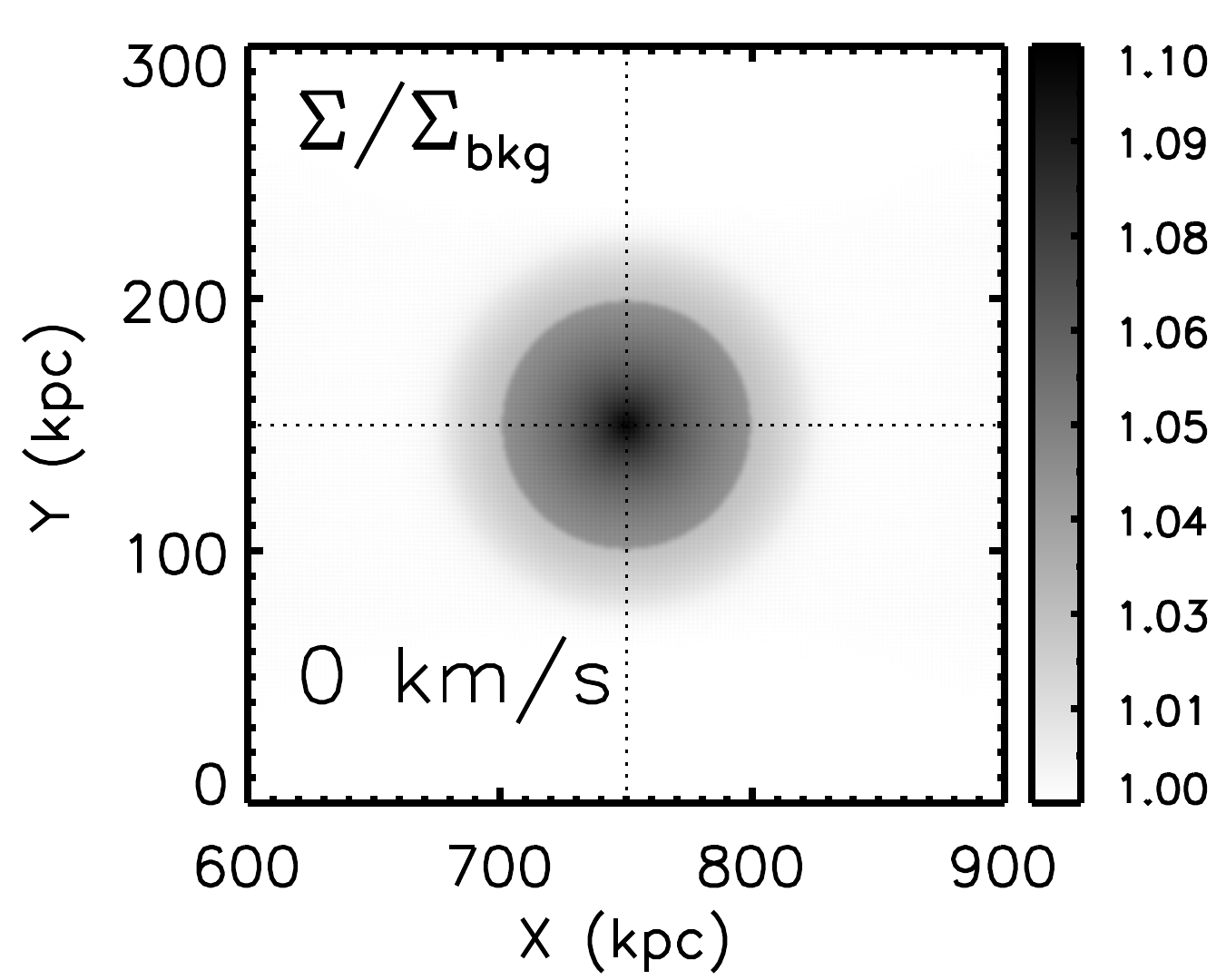}\\ 
  \includegraphics[width=0.499\columnwidth]{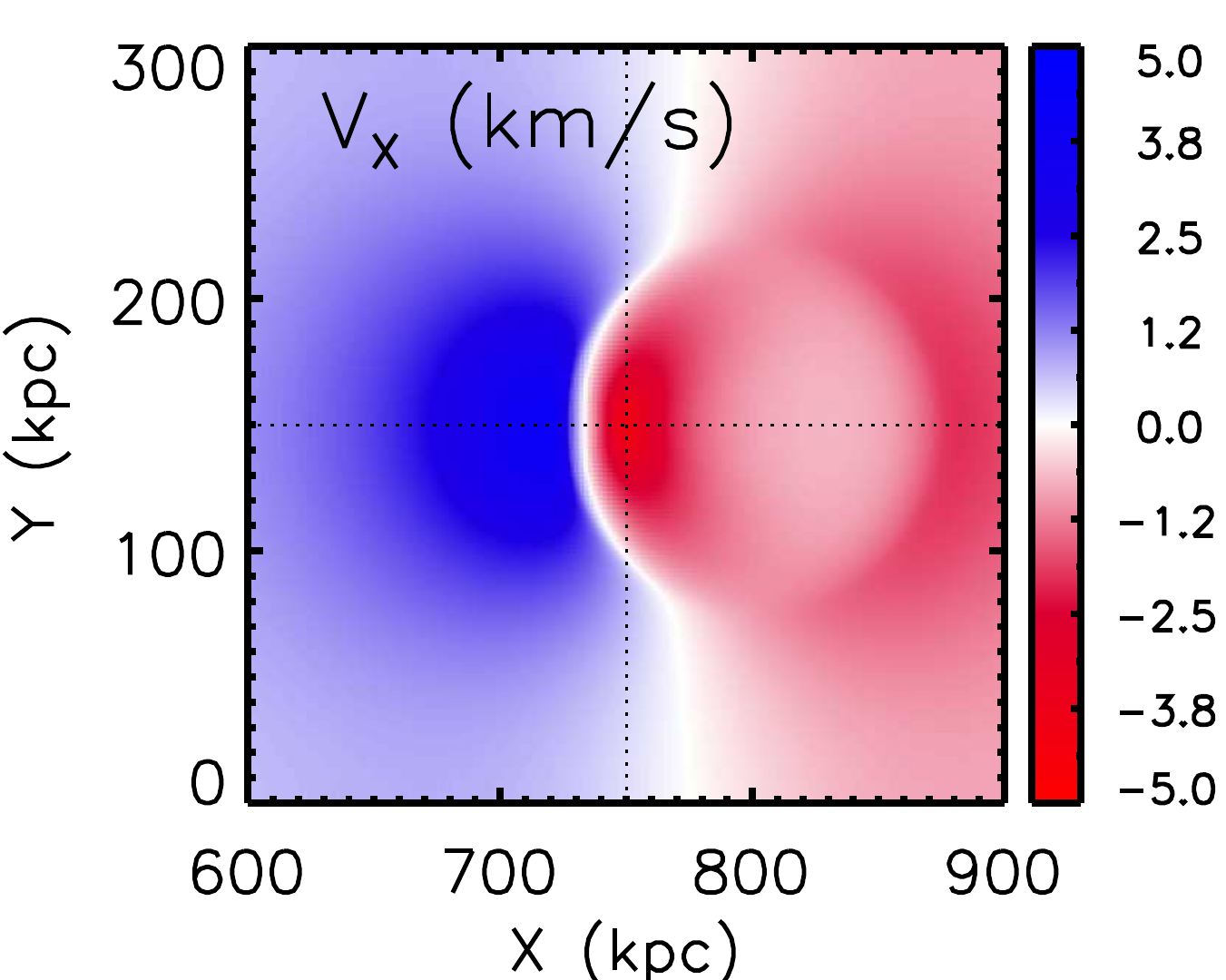}
  \includegraphics[width=0.499\columnwidth]{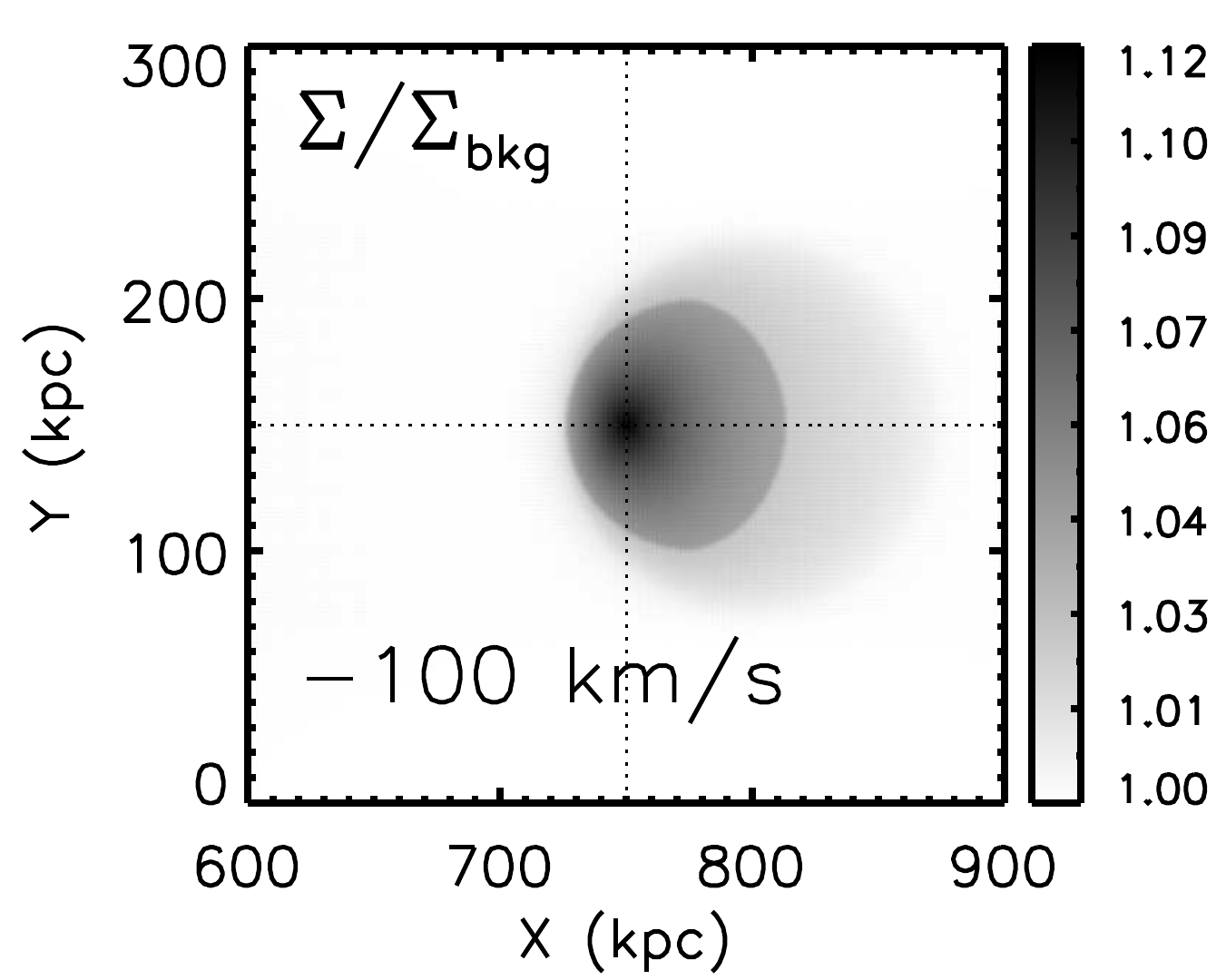}\\ 
  \includegraphics[width=0.499\columnwidth]{FIGS/M200_1E10_nIGM_m4p0_TIGM_6p0_vx_200_vx_rho_weighted_image_0050.pdf}
  \includegraphics[width=0.499\columnwidth]{FIGS/M200_1E10_nIGM_m4p0_TIGM_6p0_vx_200_sigma_image_0050.pdf}\\
  \includegraphics[width=0.499\columnwidth]{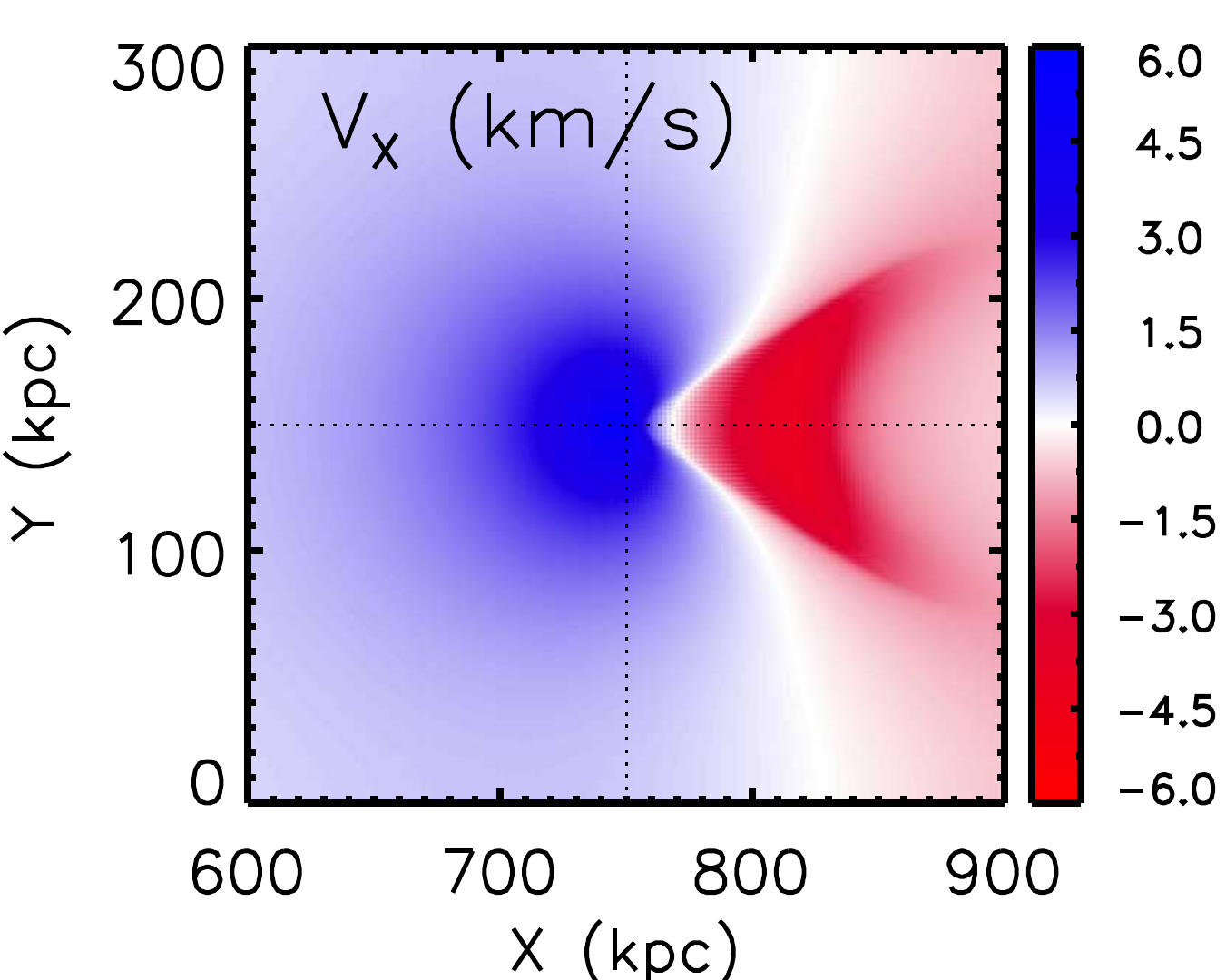}
  \includegraphics[width=0.499\columnwidth]{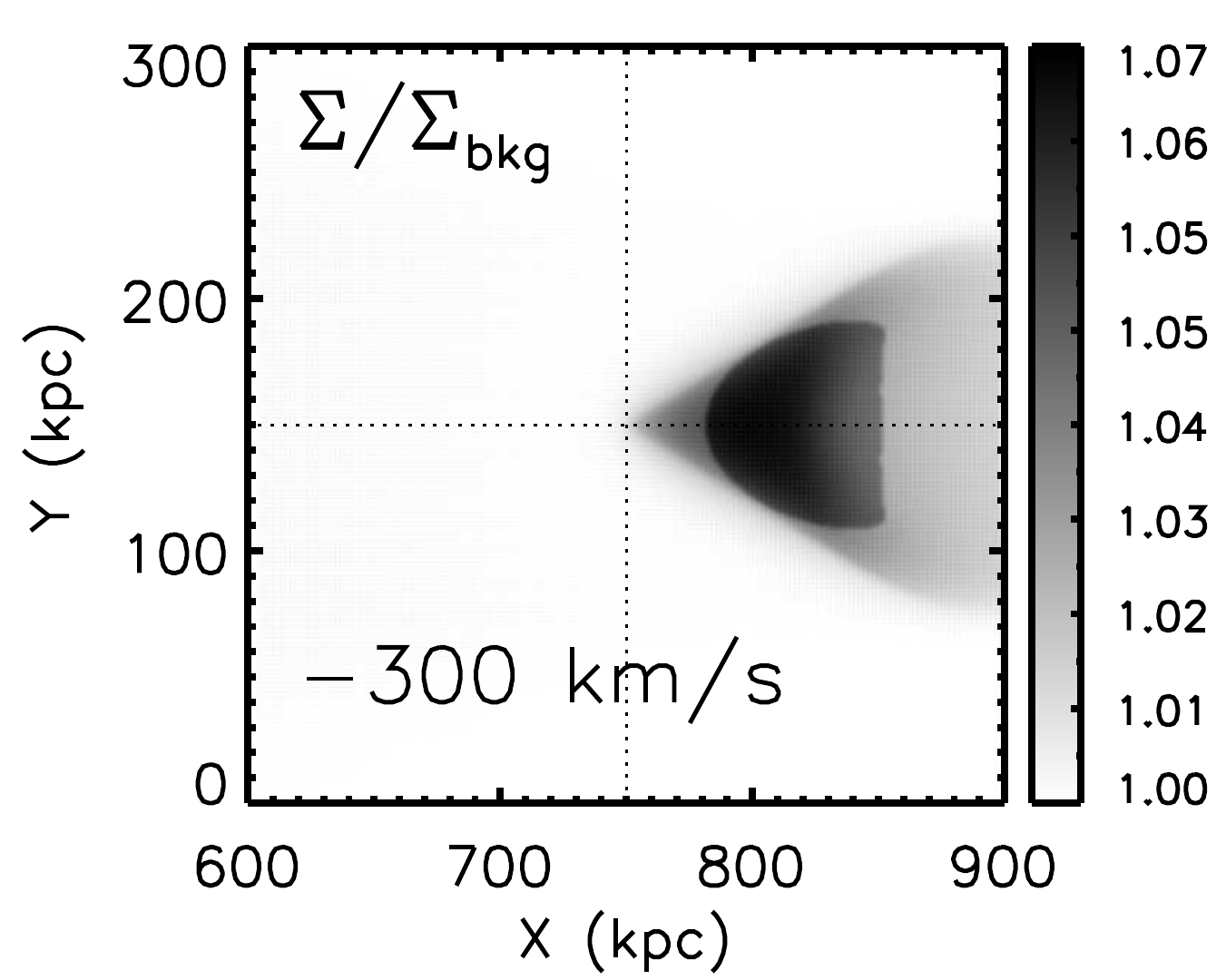}\\
  \includegraphics[width=0.47\columnwidth]{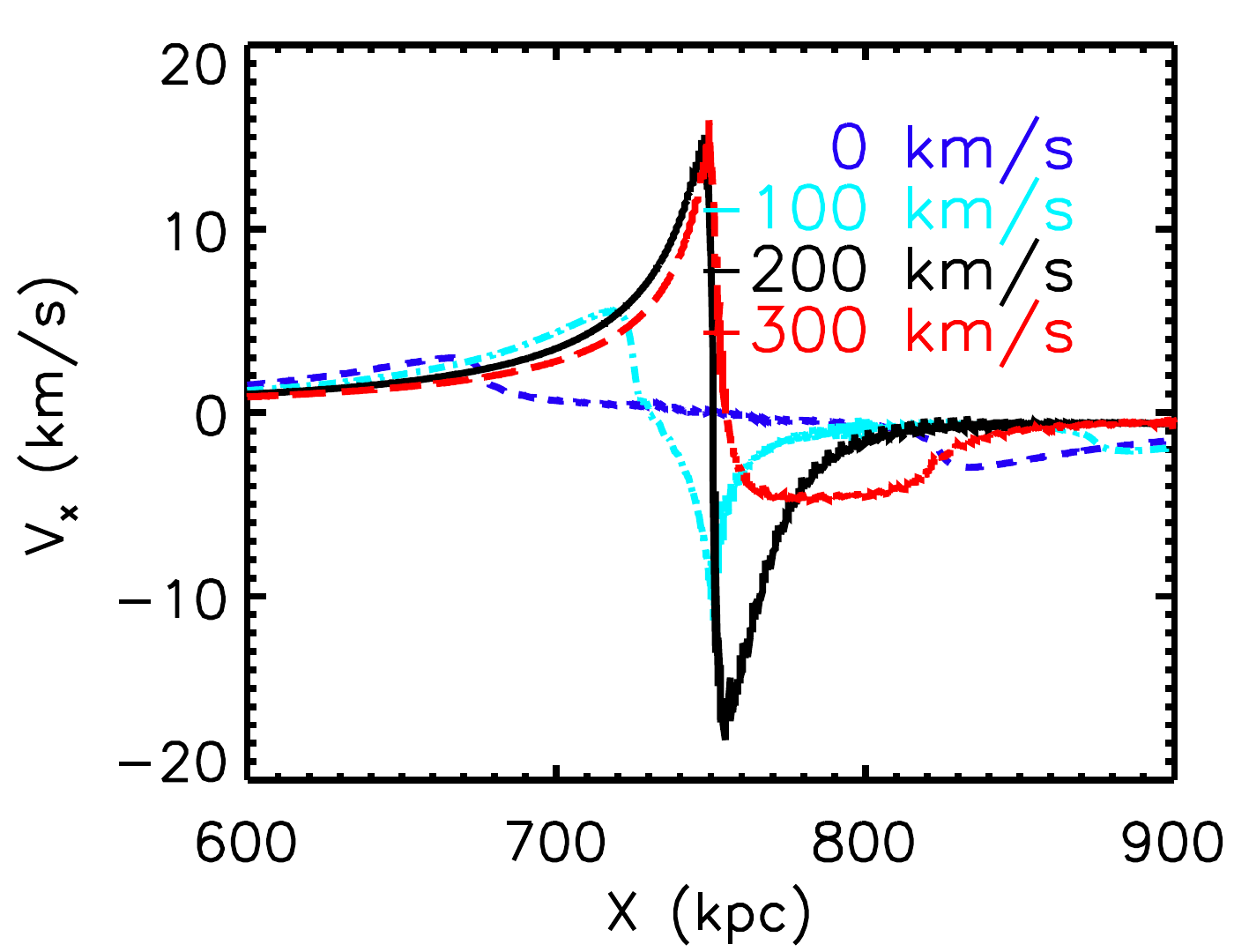} \ \
  \includegraphics[width=0.47\columnwidth]{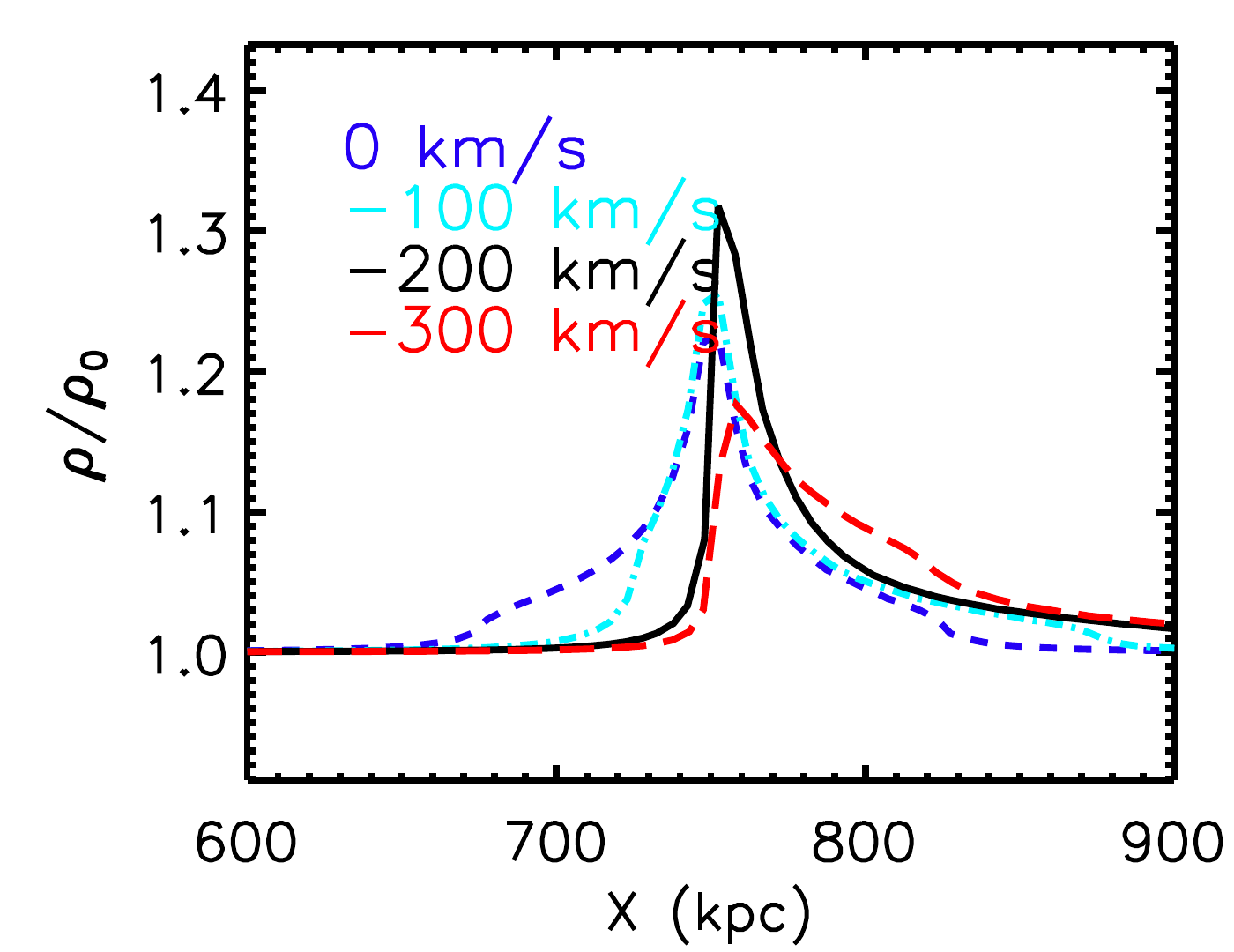}
  \caption{The impact of varying the subhalo velocity on the compression of the CGM.  The rows with images ($x$-component of the velocity in the left column and surface mass density in the right column) are ordered according to increasing subhalo (magnitude of) velocity from top to bottom, with 0 km/s, -100 km/s, -200 km/s, and -300 km/s.  The bottom row shows 1-D profiles of the $x$ component of the velocity and the density along the $x$-axis within a narrow cylinder of width 2.5 kpc (i.e., $r_{yz} < 2.5$ kpc).}
  \label{fig:vary_velocity}
\end{figure}

\subsection{Varying the relative velocity}

As subhaloes orbit within the main host halo, their velocity with respect to the CGM changes with time.  Typically, the velocity will be of order the circular velocity of the host, but it can vary from being nearly at rest (e.g., at apocentre of a nearly radial orbit) to being up to a few times the circular velocity (at pericentre).  It is therefore of interest to examine how the velocity of the subhalo with respect to the CGM impacts the results. 

In Fig.~\ref{fig:vary_velocity} we show the resulting velocity and surface mass density maps (edge-on configuration) when increasing the magnitude of the subhalo velocity from 0 km/s (top) to -300 km/s (bottom row of images) in steps of -100 km/s.  The fiducial case corresponds to -200 km/s (third row down).  For all cases the mass of the subhalo is $10^{10}$ M$_\odot$ and the temperature of the CGM is $10^{6}$ K.   

For the at rest case (0 km/s), as expected a circularly-symmetric distribution results.  Increasing the magnitude of the subhalo velocity steadily shifts the enhanced region downstream and compresses it into a more conical morphology.  The amplitude of the effect on the density varies with the subhalo velocity but not greatly, changing only a factor of two across the four cases.

To get a more quantitative sense of the variations, in the bottom row of Fig.~\ref{fig:vary_velocity} we show 1-D profiles (along the $x$ axis) of the $x$ component of the velocity and the gas density.  These profiles were constructed by selecting particles within a narrow cylinder of width 2.5 kpc along the $x$ axis.

The profiles show fairly complex behaviours.  For the at rest case, the gas is nearly hydrostatic within the very inner regions and shows only a gentle infall velocity on the outskirts.  Increasing the magnitude of the velocity to -100 km/s means that the gas is unable to establish a hydrostatic configuration centred on the subhalo which, in turn, reduces pressure near the subhalo centre and consequently upstream gas can infall slightly faster.  Increasing to -200 km/s and -300 km/s shifts the enhanced regions further downstream, so upstream gas is able to infall towards the subhalo up to a maximum velocity of $\approx15$ km/s.  (Note that the 1-D velocities are higher than inferred from the 2-D maps, which is due to averaging along the line of sight.)  The downstream velocity configuration is quite different between the -200 km/s and -300 km/s cases, though, with the latter showing only a minimal velocity towards the subhalo.  This is plausibly due to the shorter time the downstream gas in the -300 km/s case is in close proximity to the subhalo, and is therefore not accelerated to the same degree.  

\begin{figure}
  \includegraphics[width=0.499\columnwidth]{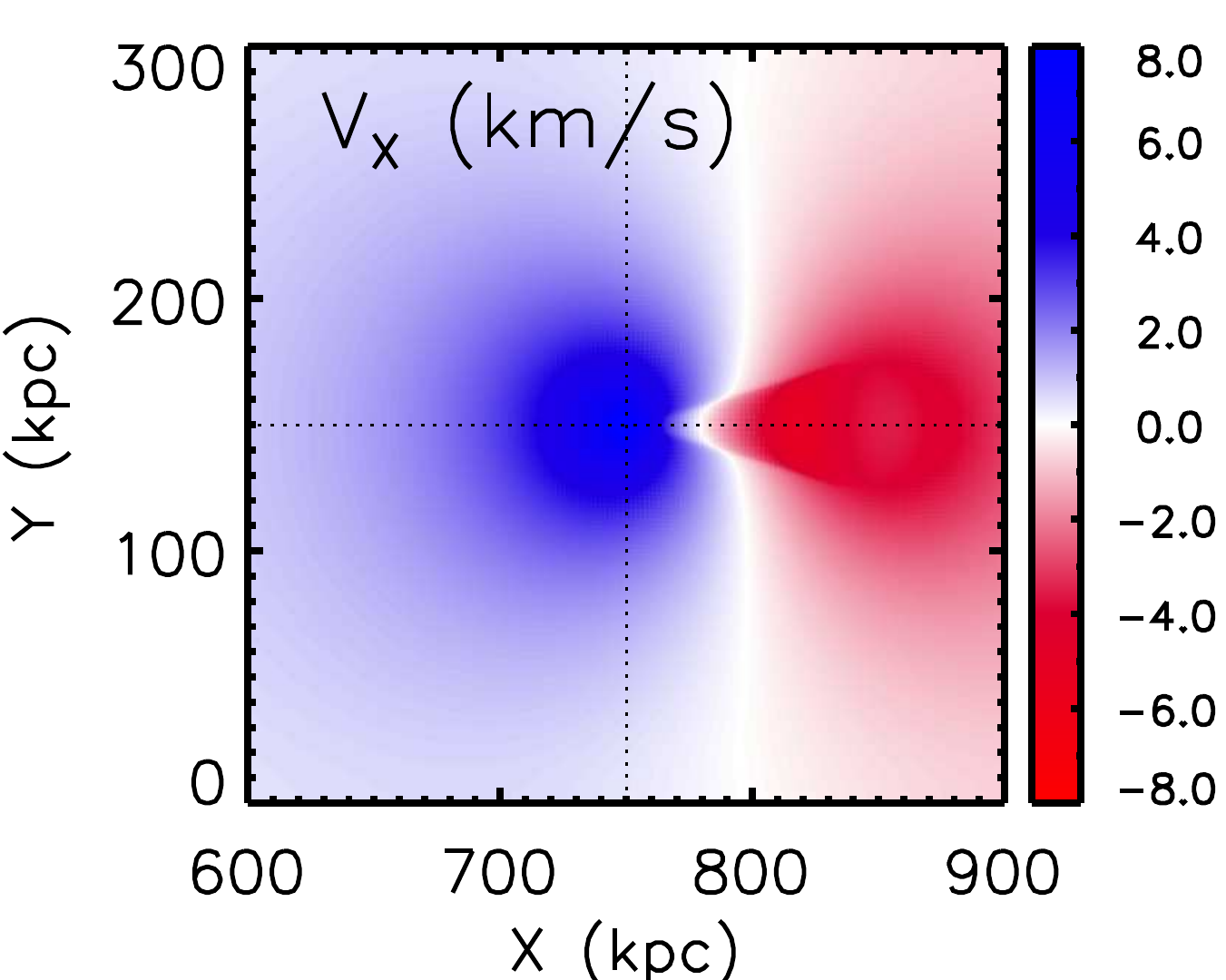}
  \includegraphics[width=0.499\columnwidth]{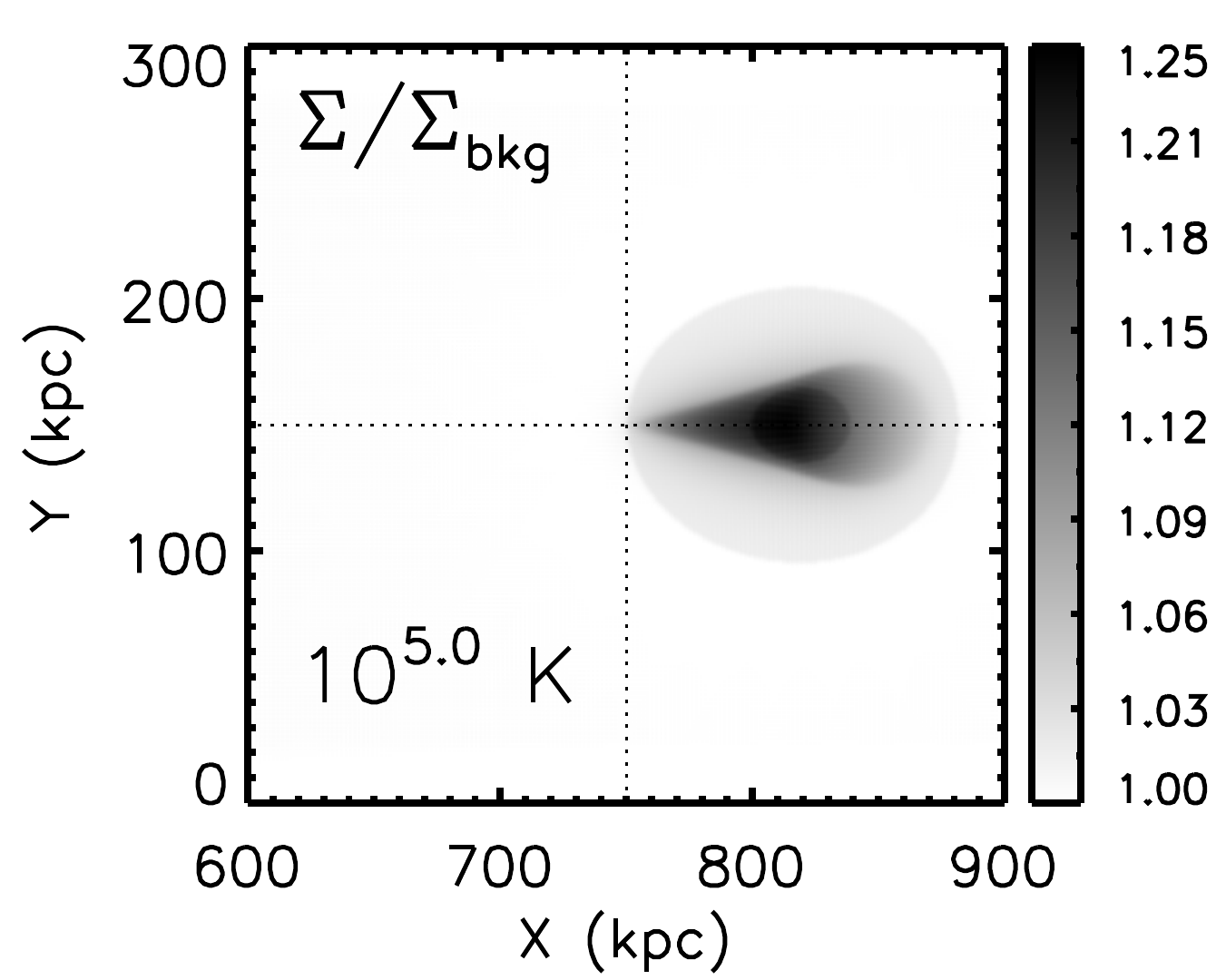}\\ 
  \includegraphics[width=0.499\columnwidth]{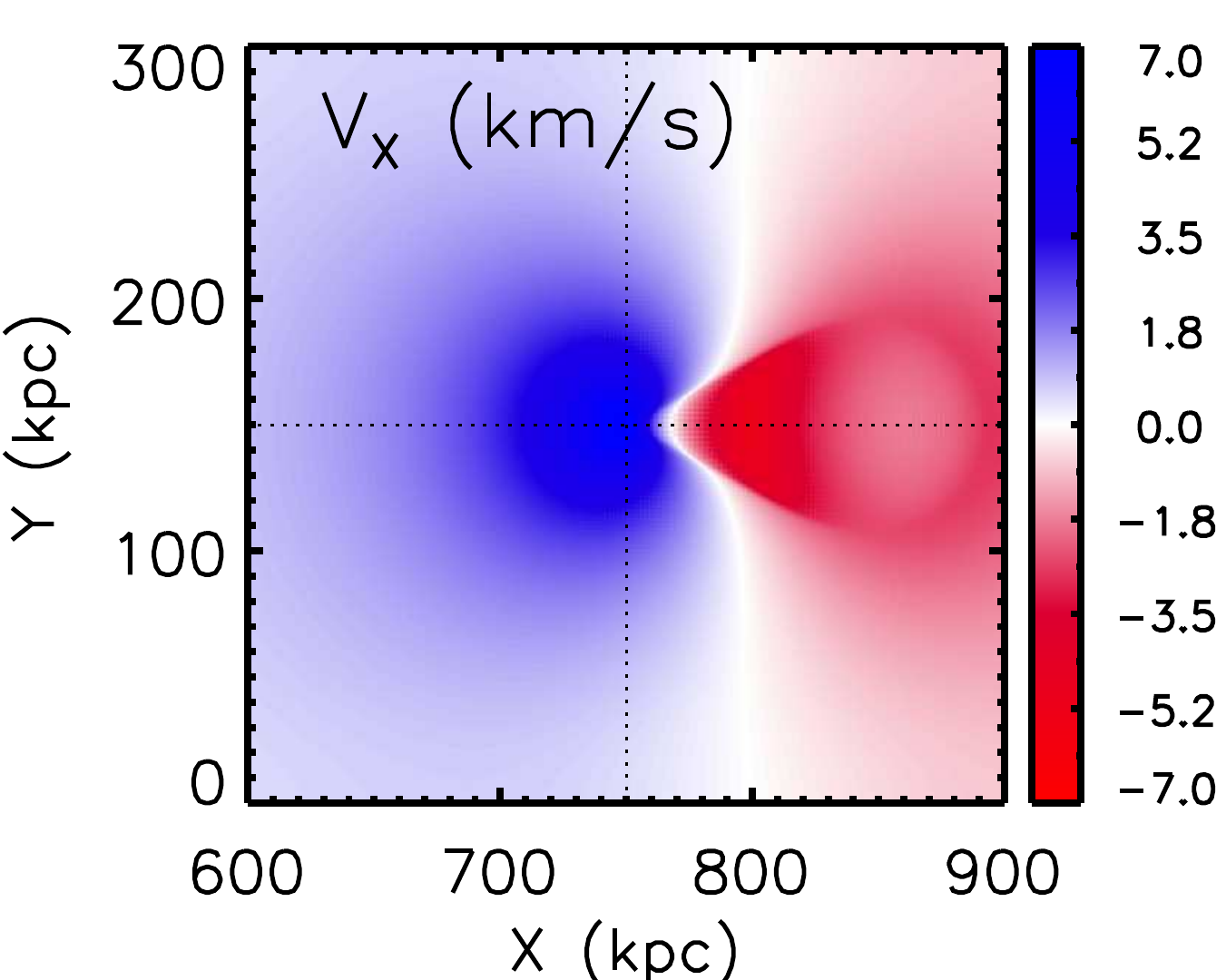}
  \includegraphics[width=0.499\columnwidth]{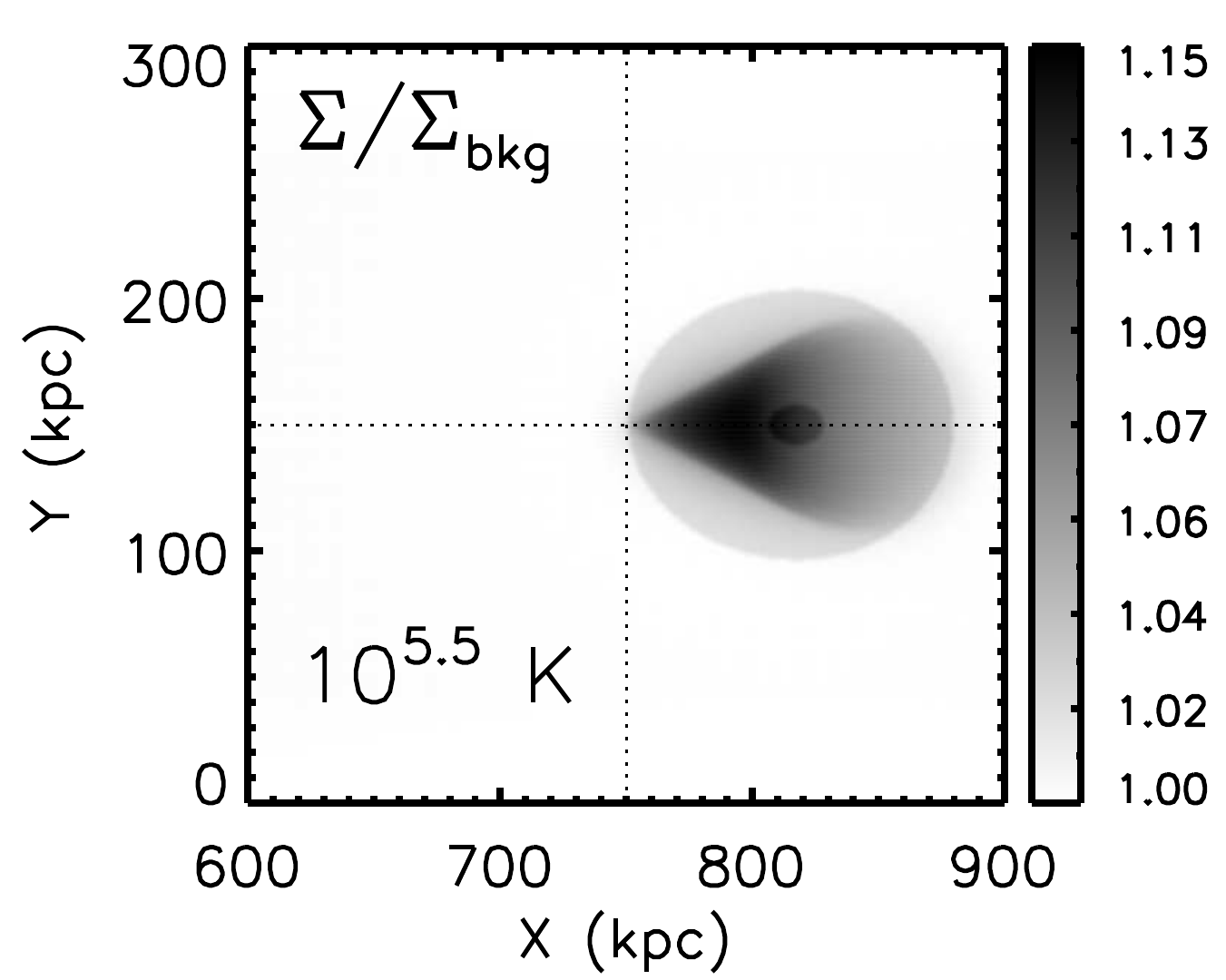}\\ 
  \includegraphics[width=0.499\columnwidth]{FIGS/M200_1E10_nIGM_m4p0_TIGM_6p0_vx_200_vx_rho_weighted_image_0050.pdf}
  \includegraphics[width=0.499\columnwidth]{FIGS/M200_1E10_nIGM_m4p0_TIGM_6p0_vx_200_sigma_image_0050.pdf}\\
  \includegraphics[width=0.499\columnwidth]{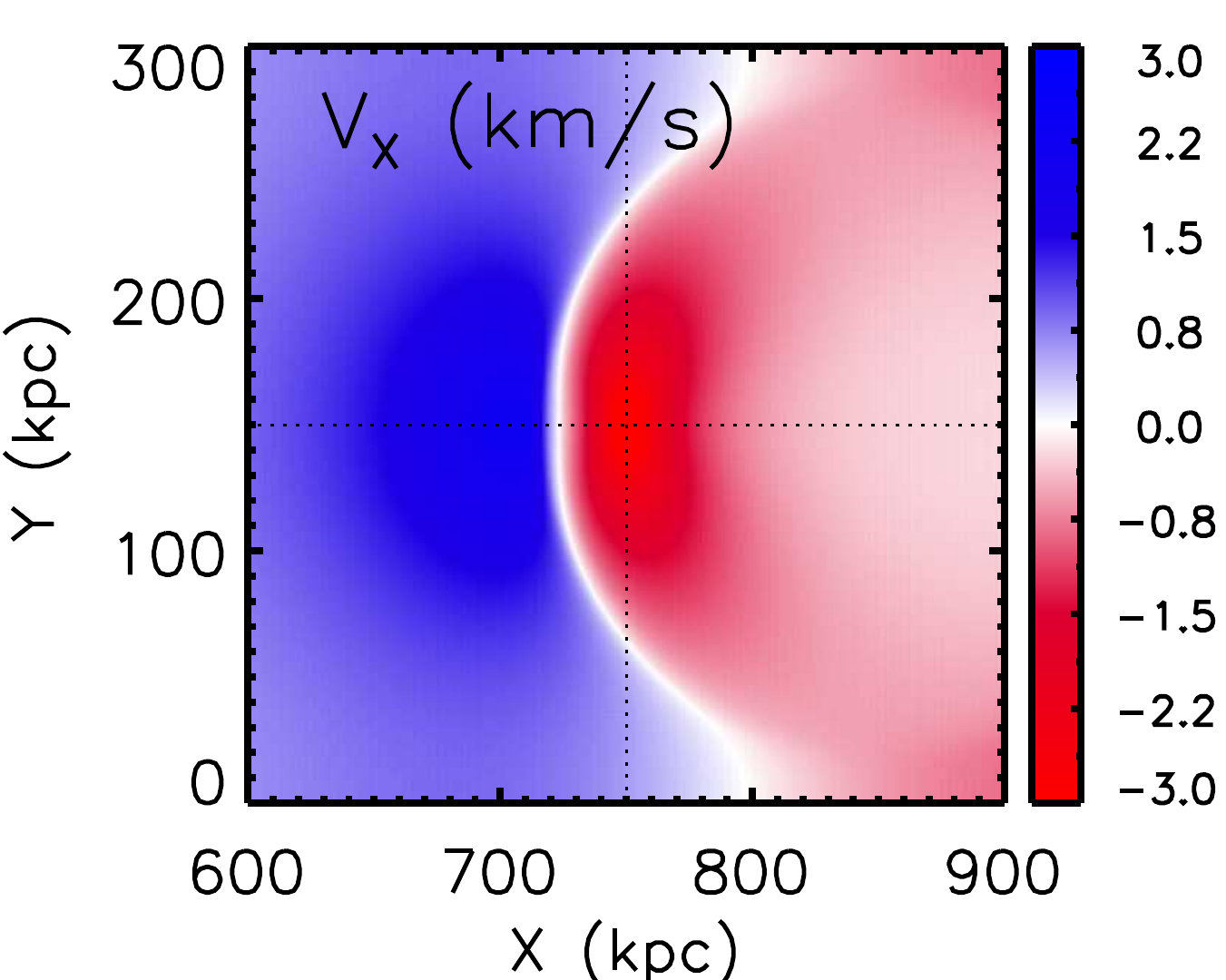}
  \includegraphics[width=0.499\columnwidth]{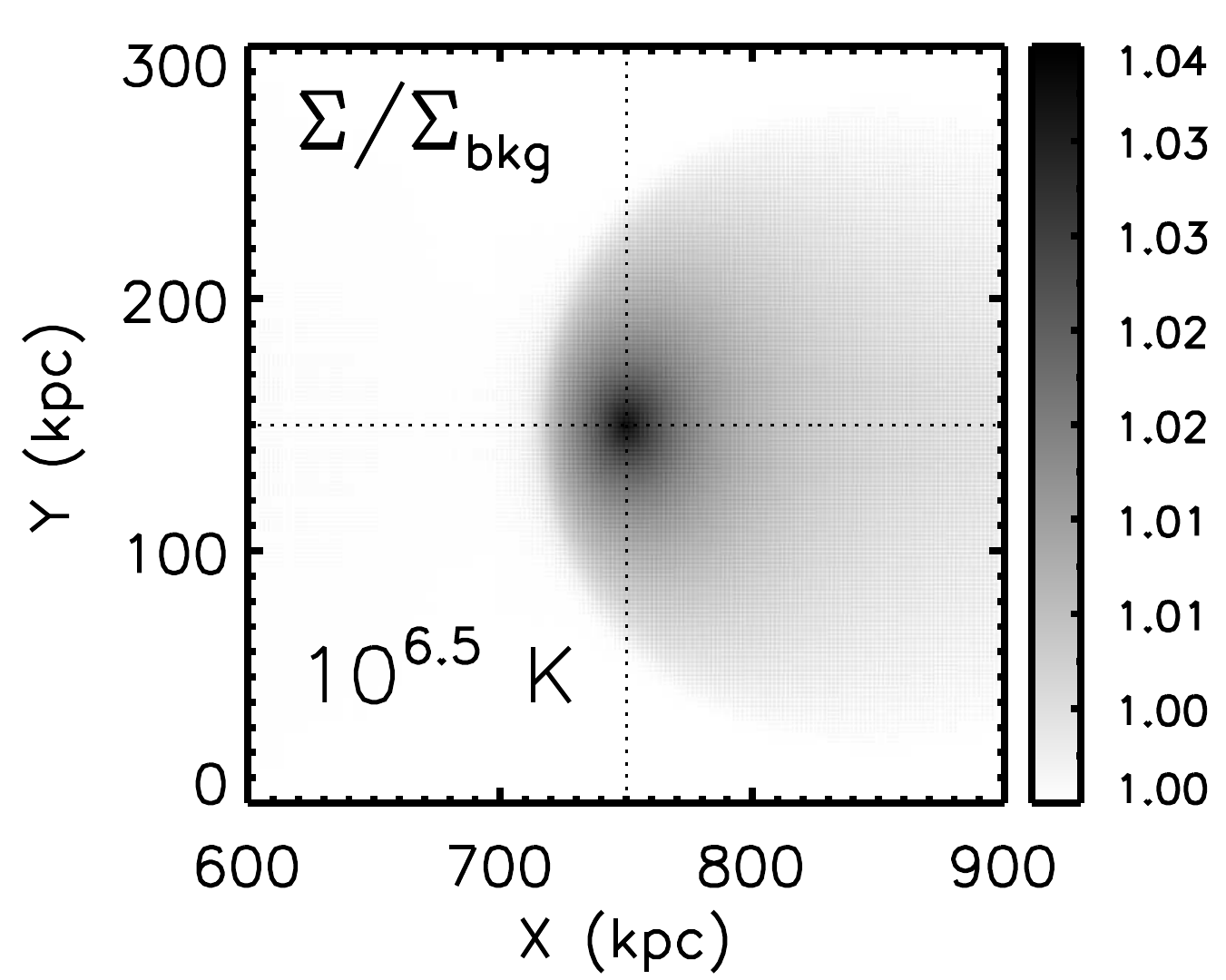}\\
  \includegraphics[width=0.47\columnwidth]{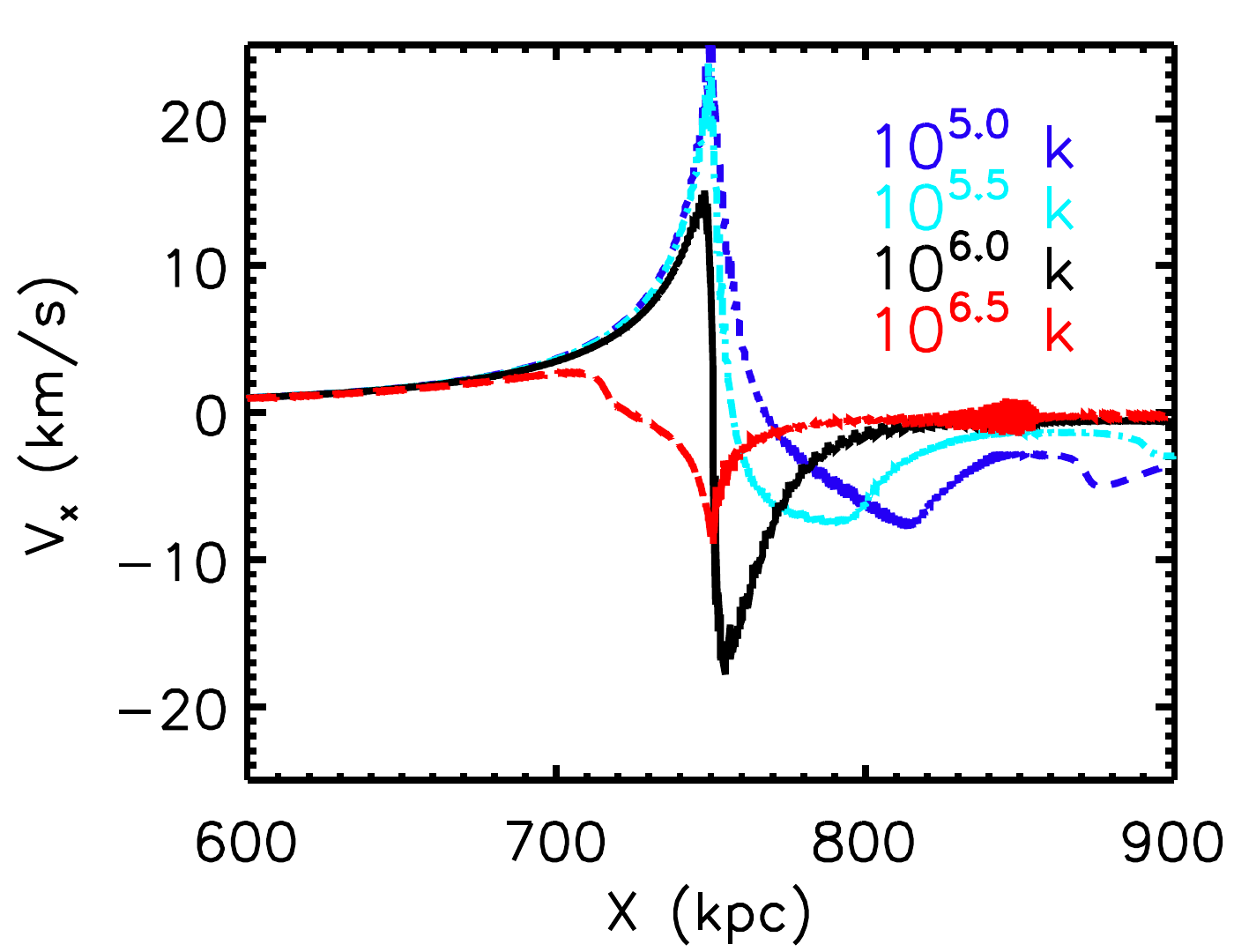} \ \
  \includegraphics[width=0.47\columnwidth]{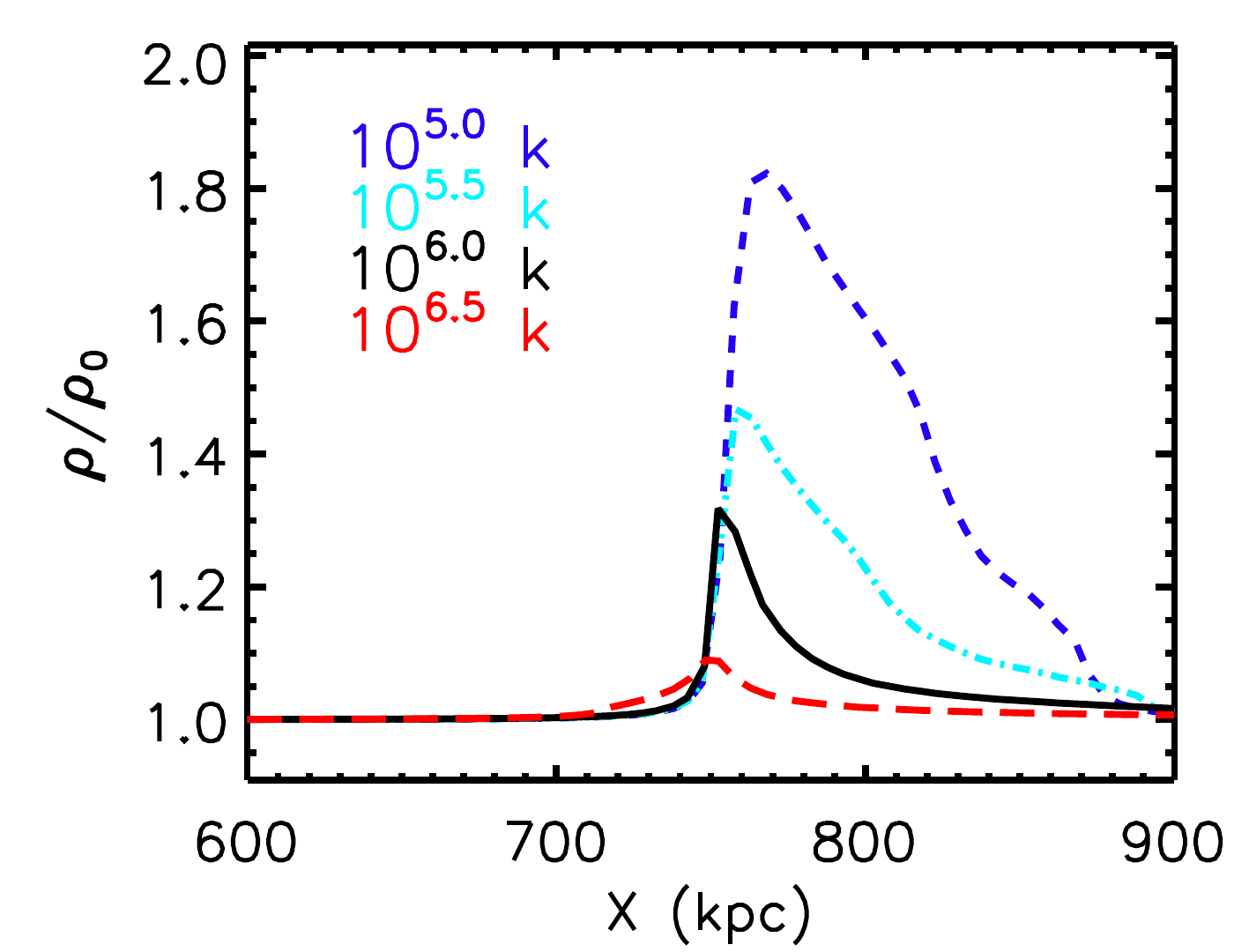}
  \caption{The impact of varying the temperature of the CGM.  The rows with images ($x$-component of the velocity in the left column and surface mass density in the right column) are ordered according to increasing CGM temperature from top to bottom, with $10^{5.0}$ K, $10^{5.5}$ K, $10^{6.0}$ K, and $10^{6.5}$ K.  The bottom row shows 1-D profiles of the $x$ component of the velocity and the 3-D density along the $x$-axis within a narrow cylinder of width 2.5 kpc (i.e., $r_{yz} < 2.5$ kpc).}  
  \label{fig:vary_temp}
\end{figure}

\subsection{Varying the CGM temperature}
\label{sec:vary_temp}

The mean temperature of the CGM is expected to be of order the virial temperature for massive haloes and therefore varies as $T_{\rm cgm} \propto M_{\rm host}^{2/3}$ \citep{white1978}.  However, even within a single host halo the CGM is a multiphase plasma characterised by a wide range of temperatures (e.g., \citealt{vandevoort2012}), which has been confirmed observationally (e.g., \citealt{tumlinson2017}).  From the analysis in Section \ref{sec:analytic}, we expect the amplitude of the effects on the CGM thermodynamic quantities to be strongly dependent on the ratio of the CGM temperature to the virial temperature (mass) of the subhalo.  As we show below, the resulting morphology is also strongly affected by this ratio.

In Fig.~\ref{fig:vary_temp} we show the resulting velocity and surface mass density maps (edge-on configuration) when increasing the CGM temperature from $10^{5.0}$ K (top) to $10^{6.5}$ K (bottom row of images) in steps of half a dex.  The fiducial case corresponds to $10^{6.0}$ K (third row down).  For all cases the mass of the subhalo is $10^{10}$ M$_\odot$ and the velocity is -200 km/s.  For reference, if we convert the CGM temperatures into isothermal sound speeds, via $c_s = (k_BT/\mu m_H)^{1/2}$, we find $c_s\approx 36.8$, $65.4$, $116.3$ and $206.9$ km/s for the four cases.

Increasing the CGM temperature has two key effects: i) the magnitude of the effect of the subhalo (i.e., the compression of the CGM) is reduced; and ii) the morphology changes such that shape becomes more symmetrical and is more centred on the subhalo.  While the first effect can be readily understood from the analytic arguments presented in Section \ref{sec:analytic}, the second effect cannot.  That is because the analytic calculation considered a CGM that was at rest with respect to the subhalo.  Here the subhalo is moving at -200 km/s with respect to the CGM for all cases and yet the morphology differs strongly between them.  

The origin of this behaviour is the due to the variation in the sound speed of the CGM.  Specifically, as we increase the temperature of the CGM we are increasing its sound speed and therefore decreasing the time over which a pressure wave can traverse the CGM.  At higher sound speeds, the gas can more rapidly adjust its configuration in response to the presence of a subhalo perturber, which is why the gas is more symmetric and centred on the subhalo in the case where $T_{\rm cgm} = 10^{6.5}$ K.  At low temperatures, the opposite is true and it is only after the subhalo has passed by that the gas is able to react, resulting in a downstream arrowhead-like configuration.

In the bottom row of Fig.~\ref{fig:vary_temp} we again show 1-D profiles (along the $x$ axis) of the $x$ component of the velocity and the gas density.  These confirm the very strong dependence of the compression of the CGM on its temperature at fixed subhalo mass.  The velocity profiles show interesting behaviour.  Specifically, for the highest temperature case the gas is able to react so quickly that there is effectively a bow wave that prevents upstream gas from falling towards the oncoming subhalo and actually drives it away at a mild velocity.

\subsection{Varying the subhalo mass}

The final variation we consider is the mass of the subhalo.  If the local gravitational compression of the CGM due to the presence of subhaloes is to be a competitive probe of dark matter physics, it will need to be sensitive to subhaloes of $10^{8-9}$ M$_\odot$.  For example, under the assumption that the dark matter particle is a thermal relic, current Lyman-alpha forest and strong lensing data constrain its mass to be at least a few keV (e.g., \citealt{viel2013,irsic2017,garzilli2019,hsueh2020}).  Simulations of warm dark matter with a particle mass at this level show that the subhalo mass function is significantly affected only below $10^{10}$ M$_\odot$ (see, e.g., fig.~2 of \citealt{stafford2020}).  This suggests that these methods are probing the mass function down to this regime but not significantly further (otherwise the constraints on the particle mass would be stronger).  It is therefore of interest to see whether detailed observations of the CGM could allow us to access lower subhalo masses.  The further we can probe down the subhalo mass function, the stronger the constraints that can be placed on the nature of dark matter.

\begin{figure}
    \includegraphics[width=0.499\columnwidth]{FIGS/M200_1E10_nIGM_m4p0_TIGM_6p0_vx_200_sigma_image_0050.pdf}
    \includegraphics[width=0.499\columnwidth]{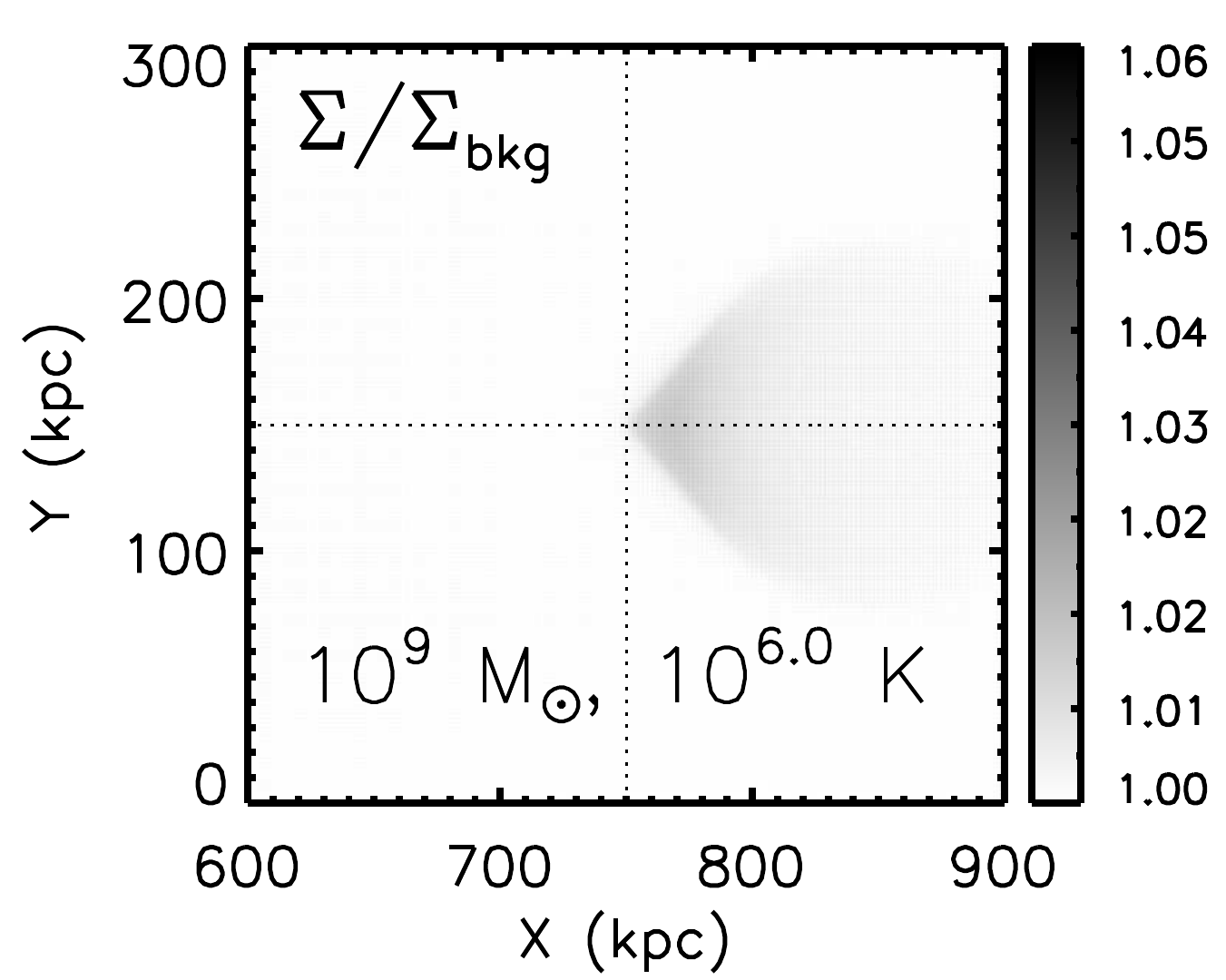}\\
    \includegraphics[width=0.499\columnwidth]{FIGS/M200_1E10_nIGM_m4p0_TIGM_5p0_vx_200_sigma_image_0050.pdf}
    \includegraphics[width=0.499\columnwidth]{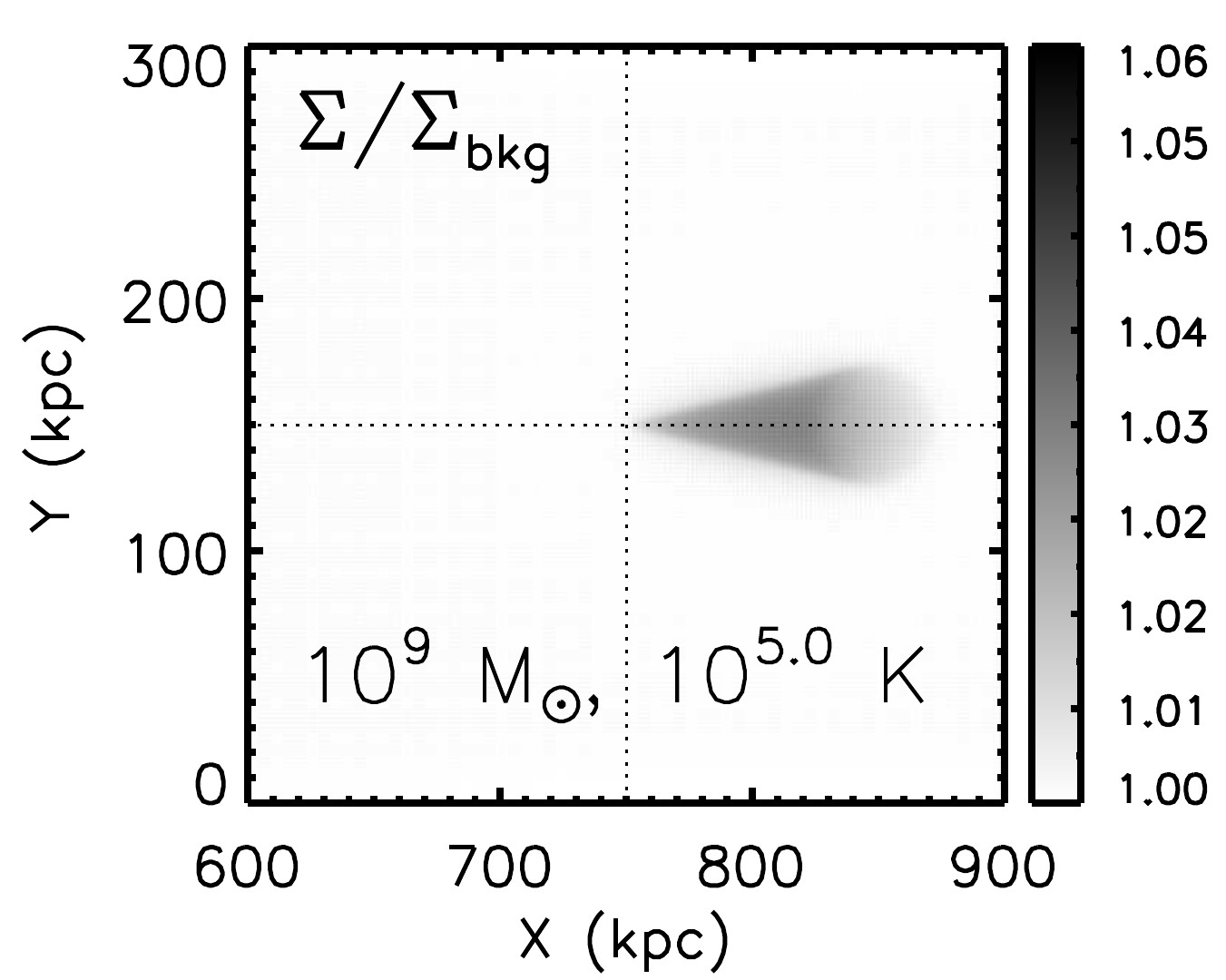}    
    \caption{The impact of varying the subhalo mass on the surface mass density maps for an edge-on view.  The top row compares the results for a $10^{10}$ M$_\odot$ subhalo (left) with that of a $10^9$ M$_\odot$ subhalo (right), both for the fiducial CGM temperature of $10^6$ K and a subhalo velocity of -200 km/s.  The bottom row compares these two subhalo masses but for a lower CGM temperature of $10^5$ K.}
    \label{fig:vary_msub_sigma}
\end{figure}

In Fig.~\ref{fig:vary_msub_sigma} we show the resulting surface mass density maps for the default subhalo mass of $10^{10}$ M$_\odot$ (left column) and for a subhalo of mass $10^9$ M$_\odot$ (right column).  The top row corresponds to the fiducial CGM temperature of $10^6$ K while the lower row corresponds to a cooler temperature of $10^5$ K.

Examining first the top row of Fig.~\ref{fig:vary_msub_sigma}, we find that the resulting morphologies for the compressed region are very similar for the two subhalo masses.  However, there is a large difference in the amplitude, such that the more massive subhalo more strongly compresses the CGM, as expected.  When projected through a column of 100 kpc, the lower subhalo mass case produces only a few percent enhancement in the surface mass density relative to the background.  Projected through a shorter column, the enhancement would obviously be somewhat larger.

From Section \ref{sec:vary_temp}, we have seen that the temperature of the CGM is hugely important.  Examining a cooler phase of the CGM, $10^5$ K as in the bottom row of Fig.~\ref{fig:vary_msub_sigma}, we see a stronger sensitivity to both subhalo masses.  The reduced sound speed of the CGM also affects the morphology, as previously discussed.

Thus, to probe the lower mass end of the halo mass function, CGM observations  that target the cooler phase are most efficient.  For example, O VI absorption line observations are particularly effective at picking out $10^5$ K gas in Milky Way-mass haloes (e.g., \citealt{oppenheimer2016}).  A potential caveat is that metal-enriched gas cools quite quickly through this temperature regime.  On the other hand, the ubiquitous detection of relatively large quantities of O VI around normal star-forming galaxies with the Cosmic Origins Spectrograph onboard HST \citep{tumlinson2011} suggests that there must be a fairly continual supply of enriched gas at this temperature, either through accretion from the intergalactic medium, cooling of higher temperature gas, or heating of cooler gas due to feedback.

\begin{figure}
    \includegraphics[width=0.499\columnwidth]{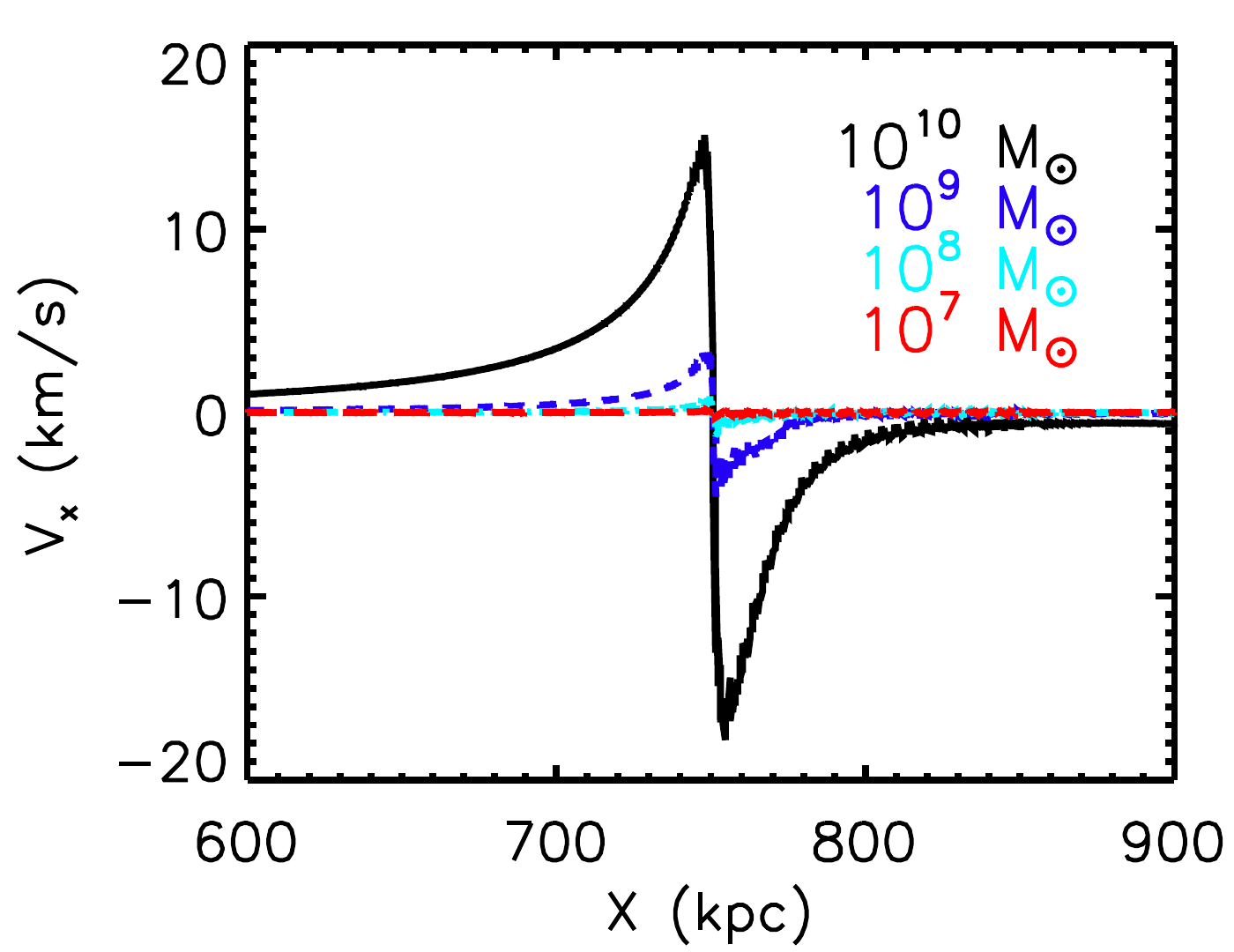}
    \includegraphics[width=0.499\columnwidth]{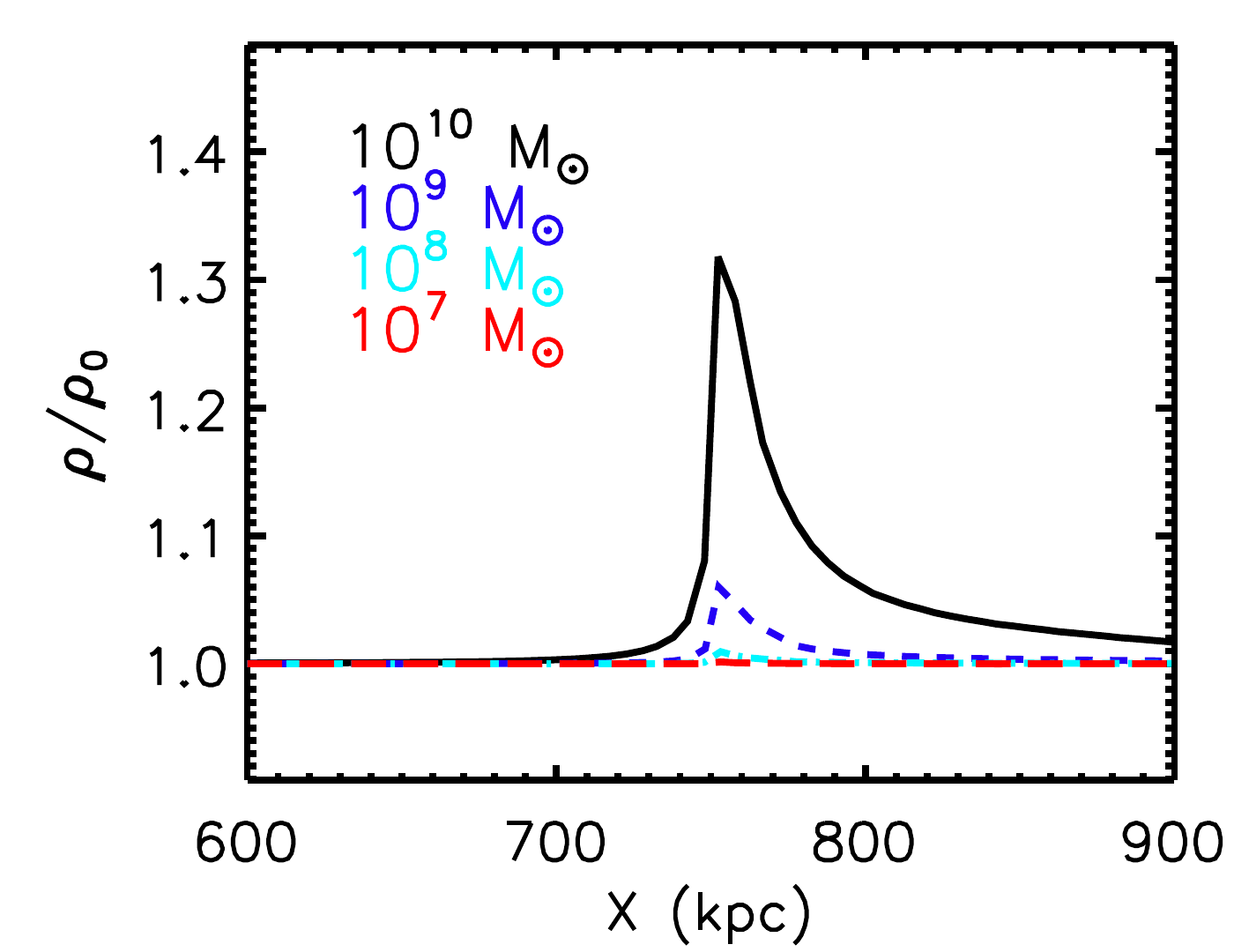}\\
    \includegraphics[width=0.499\columnwidth]{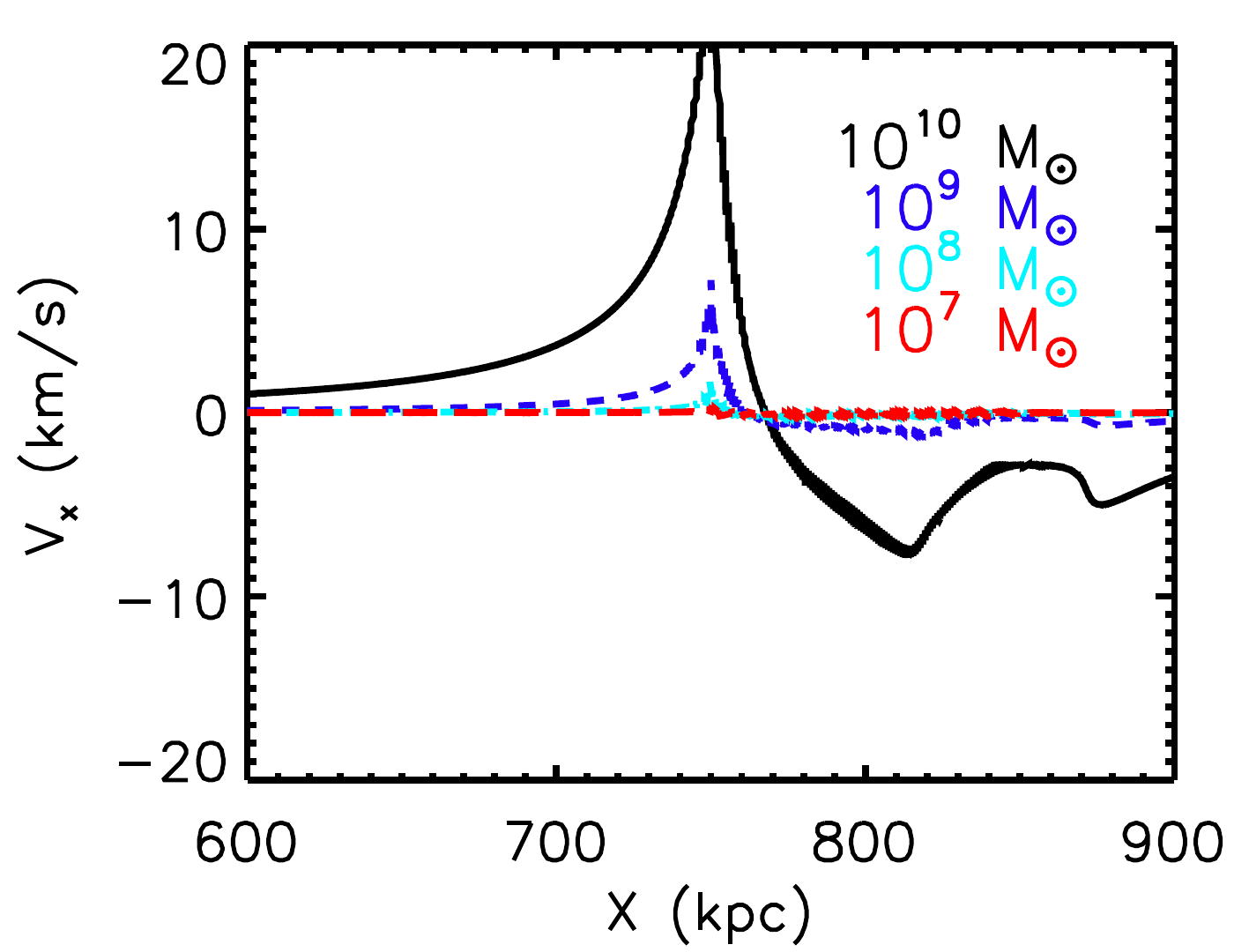}
    \includegraphics[width=0.499\columnwidth]{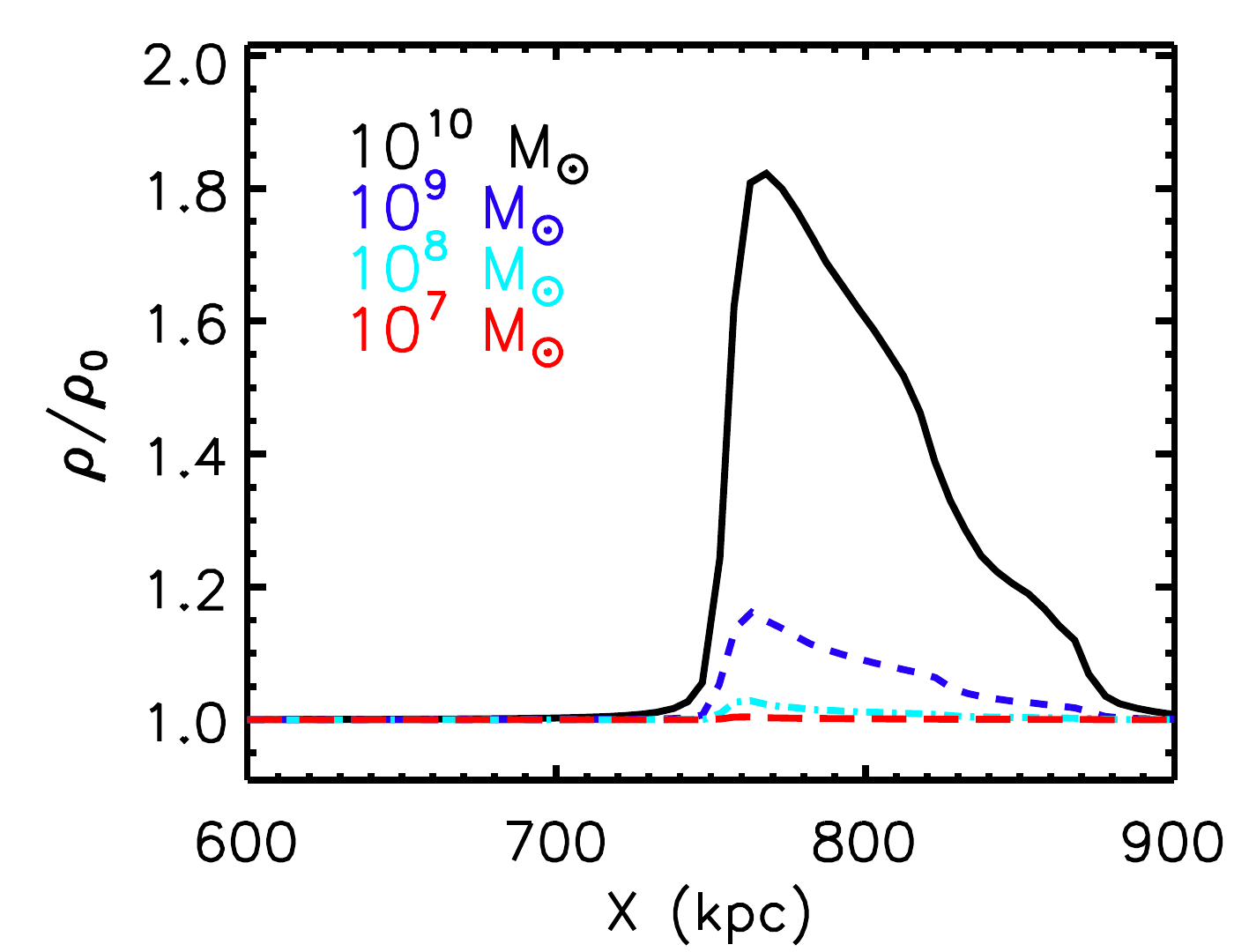}    
    \caption{1-D profiles of the $x$ component of the velocity (left column) and the 3-D density (right column) along the $x$-axis within a narrow cylinder of width 2.5 kpc (i.e., $r_{yz} < 2.5$ kpc).  Different coloured curves correspond to different subhalo masses.  The top row corresponds to the fiducial CGM temperature of $10^6$ K, while the bottom row corresponds to a lower CGM temperature of $10^5$ K.}
    \label{fig:vary_msub_profs}
\end{figure}

In Fig.~\ref{fig:vary_msub_profs} we show 1-D profiles (along the $x$ axis) of the $x$ component of the velocity (left column) and the gas density (right column) for subhaloes of varying mass (different line colours) for a CGM temperature of $10^6$ K (top row) and $10^5$ K (bottom row).  Here we consider the impact of subhaloes down to $10^7$ M$_\odot$.

Whilst examining the cooler phases of the CGM significantly boosts the signal, probing down to halo masses of $10^8$ M$_\odot$ (and lower) will clearly be a challenge.  However, even if the signal cannot be detected on an individual subhalo basis, it may still be detectable in a statistical sense.  For example, cosmic shear studies do not examine individual peaks and troughs in the lensing field, but instead compute the alignment of galaxy shapes as a function of scale averaged over a large fraction of the sky.  Analogously, one can envisage measuring the pressure, density, or temperature fluctuation power spectra of the CGM and using this as essentially a tracer for the subhalo power spectrum.  Aside from measuring these quantities, identifying the fluctuations that are driven by subhalo compression, as opposed to, for example, feedback-driven turbulence, will be a challenge that requires detailed simulations to address.

\begin{figure*}
    \includegraphics[width=\columnwidth]{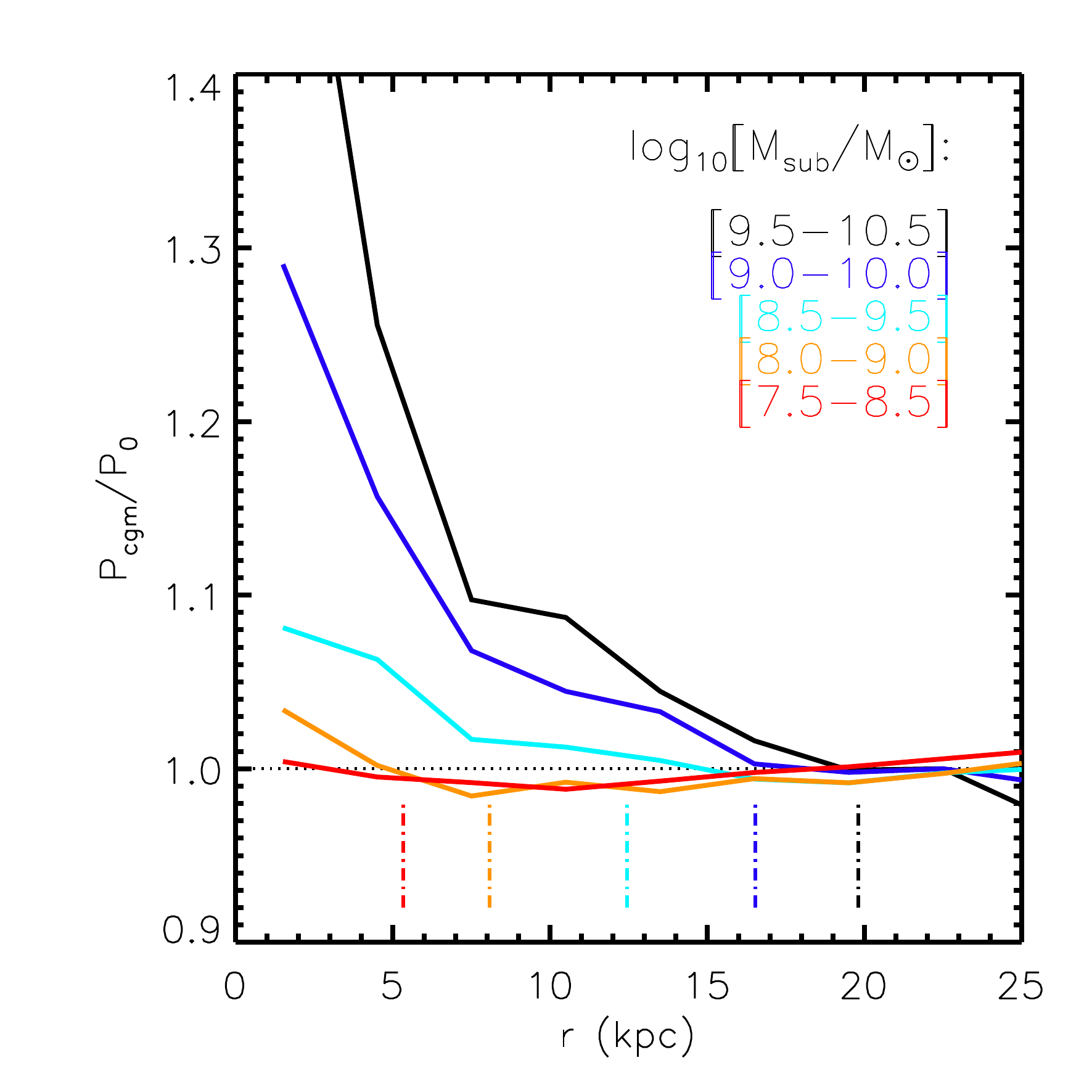}
    \includegraphics[width=\columnwidth]{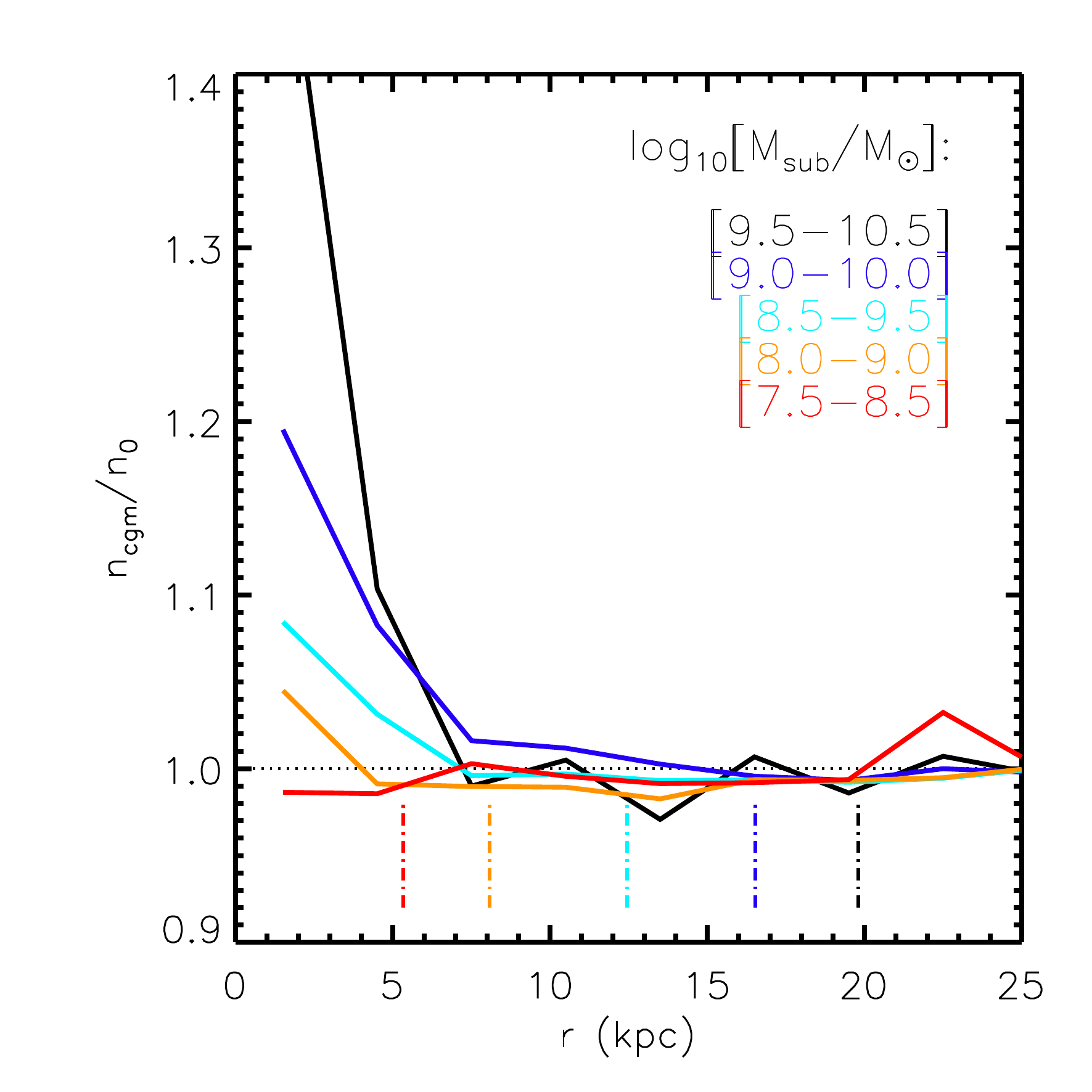}
    \caption{Stacked CGM pressure (left) and density (right) profiles centred on satellite subhaloes within Milky Way-mass host haloes in the \texttt{ARTEMIS} simulations.  Different coloured curves correspond to subhalo mass bins.  Only subhaloes which have {\it no} gravitationally-bound gas are included in the stacking analysis.  The vertical dashed lines correspond to the mean tidal radius for each subhalo mass bin, computed as the mean radius enclosing 90\% of the bound dark matter mass belonging to the dark subhaloes.}
    \label{fig:artemis}
\end{figure*}

\section{Cosmological hydro simulations: ARTEMIS}
\label{sec:artemis}

In this section we analyse a suite of cosmological hydrodynamical simulations that includes a realistic treatment of galaxy formation physics.  The aim is to verify whether or not the signal present in the idealised simulations is also present in a more realistic environment (albeit one that is simulated at lower resolution than can be achieved in the idealised simulations).  We first introduce the simulations and then present a stacked profile analysis.

\subsection{Simulation description}
We use the \texttt{ARTEMIS} suite of high-resolution cosmological hydrodynamical simulations of Milky Way-mass haloes (\citealt{font2020}, hereafter F20).  Full details of the simulations are provided in F20; we provide only a brief overview here.

The \texttt{ARTEMIS} simulations employ the `zoom in' technique (e.g.~\citealt{bertschinger2001}) to simulate Milky Way-analog haloes at high resolution and with hydrodynamics, within a larger box that is simulated at comparatively lower resolution and with collisionless dynamics only.  The initial conditions were generated using the \texttt{MUSIC} code\footnote{\url{https://www-n.oca.eu/ohahn/MUSIC/}} \citep{hahn2011}. Haloes were selected from a base periodic box is $25$~Mpc $h^{-1}$ on a side with $256^3$ particles. The initial conditions were generated at a starting redshift of $z=127$ using a transfer function computed using the \texttt{CAMB}\footnote{\url{https://camb.info/}} Boltzmann code \citep{lewis2000} for a flat $\Lambda$CDM WMAP9 \citep{hinshaw2013} cosmology ($\Omega_\textrm{m}=0.2793$, $\Omega_\textrm{b}=0.0463$, $h=0.70$, $\sigma_8=0.8211$, $n_s=0.972$).  The initial conditions include second order Lagrangian perturbation theory (2LPT) corrections.

The base periodic volume was run down to $z=0$ using the Gadget-3 code (last described in \citealt{springel2005}) with collisionless dynamics.  Milky Way analogs were selected based on total halo mass.  Specifically, F20 selected a volume-limited sample of haloes (i.e., all haloes) whose total mass fell in the range $8\times10^{11} < {\rm M}_{200}/{\rm M}_\odot < 2\times10^{12}$ at $z=0$.  Here, we use a subset of 42 high-resolution presented in F20.

The zoomed ICs were generated by first selecting all particles within $2 r_{200}$ of the selected haloes and tracing them back to the initial conditions of the periodic box, at $z=127$, to define the region which would be re-simulated at higher resolution and with baryons.

The base periodic run has a \texttt{MUSIC} refinement level of 8, whereas the zoom region has a maximum refinement level of $11$.  With this level of refinement, the DM particle mass is $1.17\times10^5$ M$_\odot\/h^{-1}$ and the initial baryon particle mass is $2.23\times10^4$ M$_\odot\/h^{-1}$.  A force resolution (Plummer-equivalent softening) of $125$~pc/$h^{-1}$ (which is in physical coordinates below $z=3$ and comoving coordinates at earlier times) was adopted, following \citet{power2003}.  Note that the resolution of \texttt{ARTEMIS} is similar to that of the highest resolution simulations from other groups for this mass scale, including the \texttt{APOSTLE}, \texttt{Auriga} and \texttt{FIRE-2} simulations (\citealt{sawala2016,grand2017,garrison-kimmel2018}, respectively).
The \texttt{ARTEMIS} sample is larger in terms of the number of Milky Way analogs simulated at this very high resolution, however, allowing us to draw more robust statistical conclusions.

\begin{figure*}
\centering
    \includegraphics[width=0.33\textwidth]{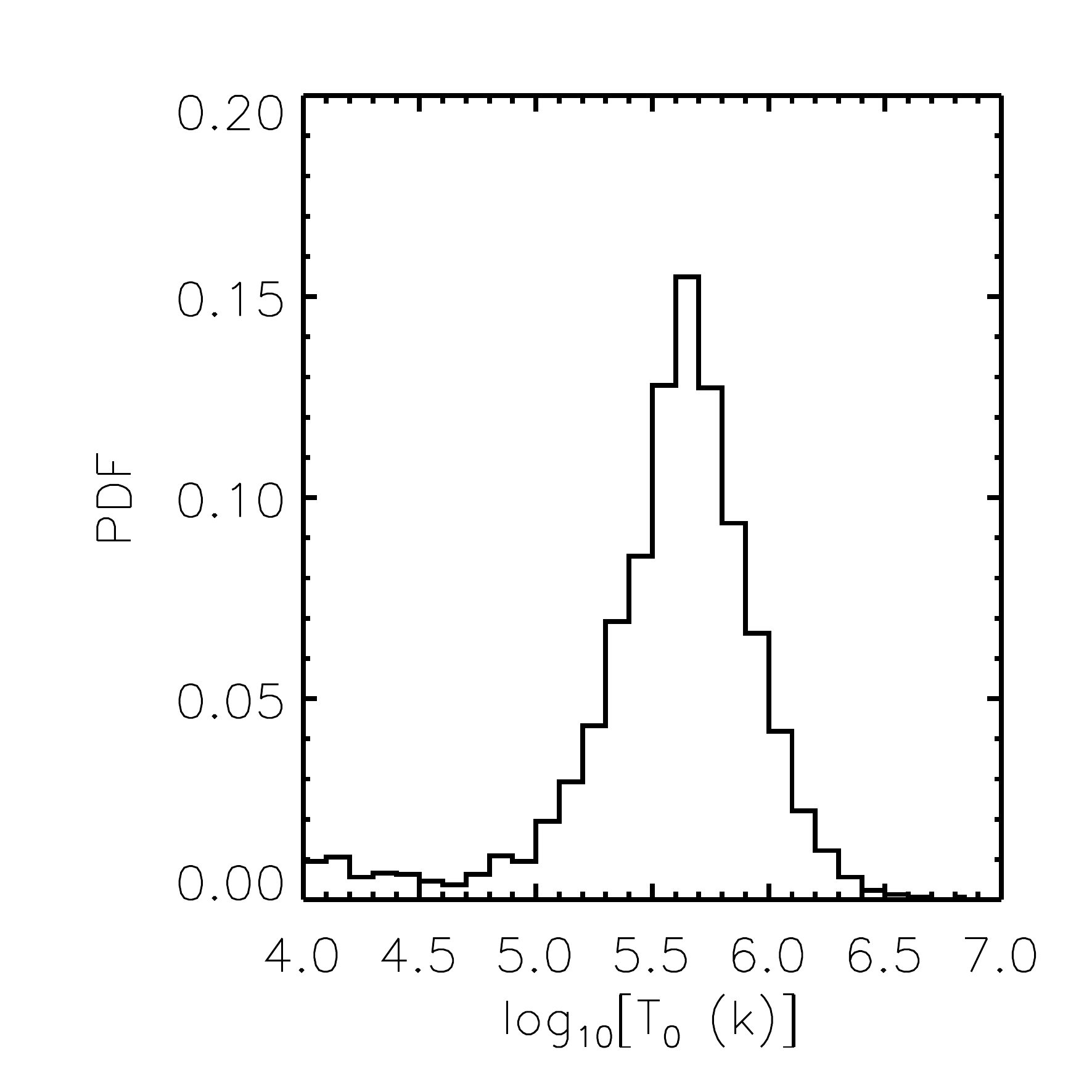}
    \includegraphics[width=0.33\textwidth]{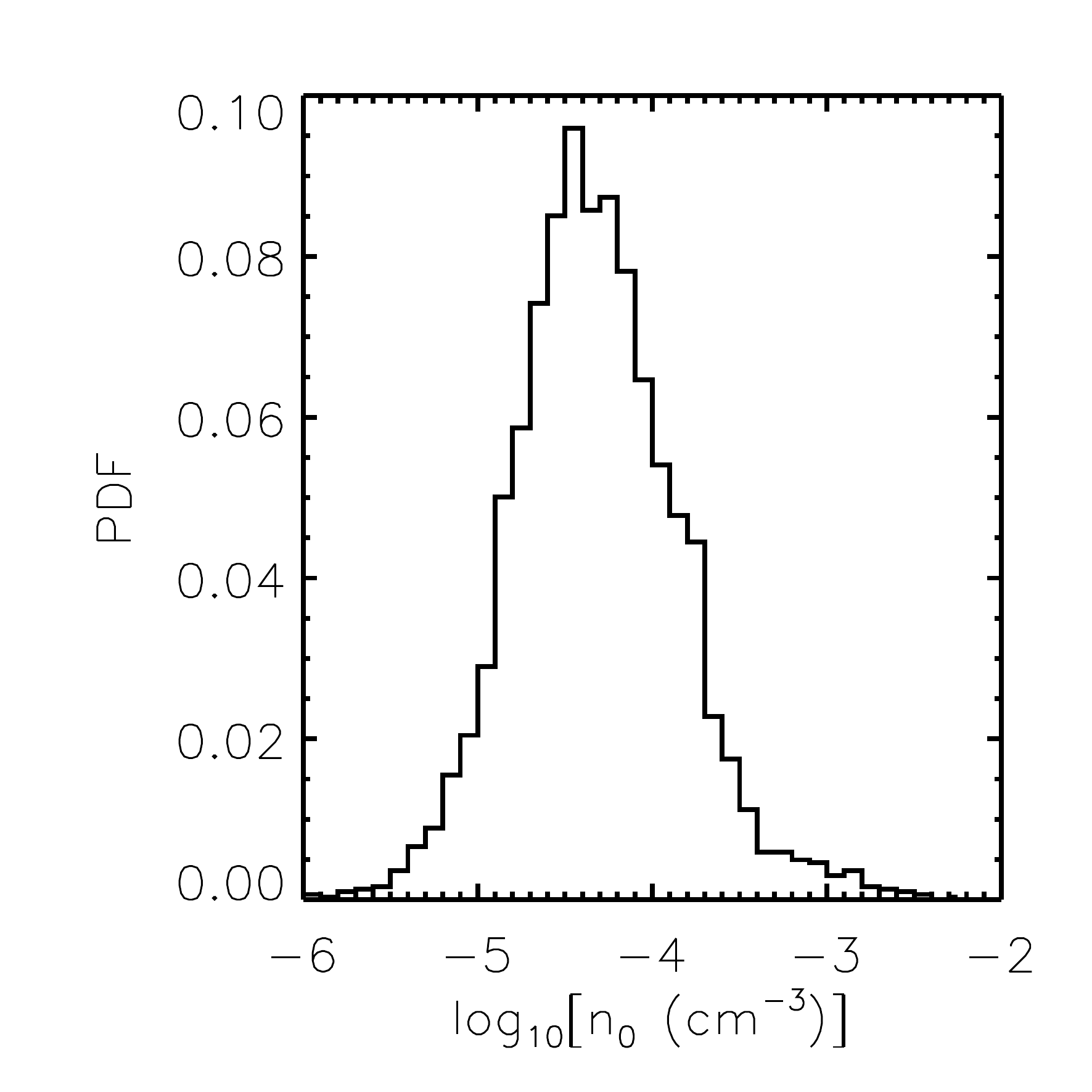}
    \includegraphics[width=0.33\textwidth]{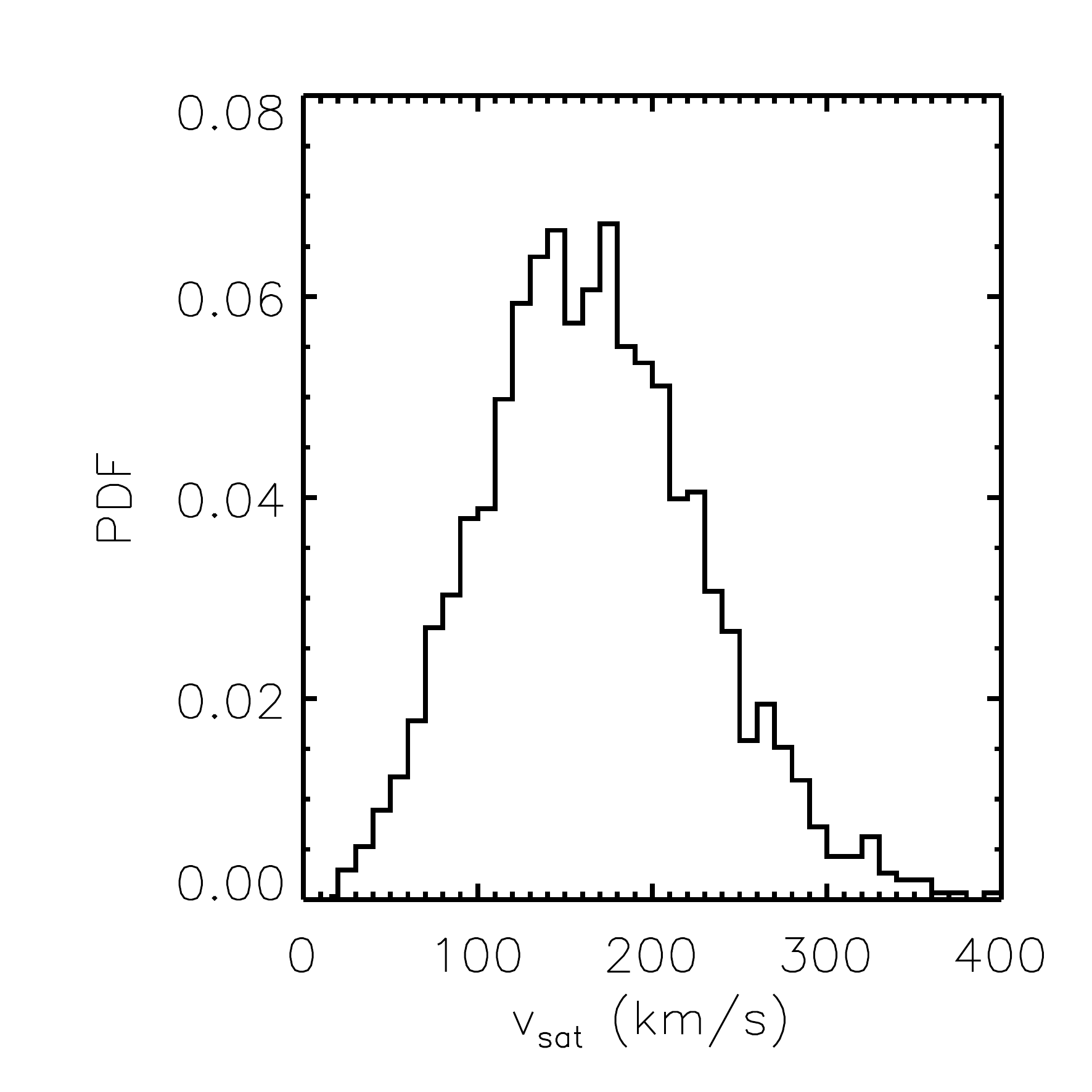}
    \caption{Probability distribution functions (PDFs) of the CGM temperature (left), CGM density (middle) and the satellite subhalo velocity with respect to the host halo (right).  The CGM temperature and density is estimated as the median value estimated in a spherical shell (with $r=25$ kpc and $\Delta r=10$ kpc) centred on each dark subhalo.}
    \label{fig:artemis_pdfs}
\end{figure*}

To carry out the hydrodynamical zoomed simulations, \texttt{ARTEMIS} uses the Gadget-3 code with an updated hydro solver and galaxy formation modelling (subgrid physics) developed for the \texttt{EAGLE} project \citep{schaye2015}.  The \texttt{EAGLE} model includes subgrid prescriptions for important processes that cannot be resolved directly in the simulations, including metal-dependent radiative cooling, star formation, stellar evolution and chemodynamics, black hole formation and growth through mergers and gas accretion, along with stellar feedback and feedback from active galactic nuclei (AGN) (see \citealt{schaye2015} and references therein).  An important consideration for galaxy formation modelling is the calibration of the feedback efficiencies.  At the scale of Milky Way-mass galaxies, stellar feedback is expected to dominate over that of AGN.  F20 adjusted the parameter values of the stellar feedback model in the \texttt{EAGLE} code to reproduce the amplitude of the observed galaxy stellar mass--halo mass relation at the Milky Way halo-mass scale.  While the simulations were not calibrated on other aspects of the observed galaxy population, they nevertheless reproduce a number of key observables, including the disc size--stellar mass and star formation rate--stellar mass relations.

\subsection{Stacked CGM profiles centred on subhaloes}

We use the \textsc{SUBFIND} algorithm \citep{springel2001,dolag2009} to identify subhaloes in the \texttt{ARTEMIS} simulations.  For each zoom simulation, we select all satellite subhaloes within the Milky Way-analog friends-of-friends group for analysis.  To avoid any ambiguity as to the origin of the gas within and near the subhaloes, we select only subhaloes which, according to \texttt{SUBFIND}, have no gravitationally-bound gaseous component whatsoever.  The vast majority of the low-mass subhaloes meet this selection criterion, as they have been efficiently ram pressure stripped due to their relative motion with respect to the CGM (as well processes such as feedback and reioinization).

In Fig.~\ref{fig:artemis} we present stacked pressure and density profiles of CGM centred on subhaloes in various total mass bins.  We use the most bound particle as the subhalo centre.  We normalise the profiles by the median pressure/density value in a spherical shell between 20 and 30 kpc (which we define as $P_0$ and $n_0$, respectively), where the profiles show significant flattening and a convergence to the background (unperturbed) CGM.  When stacking the profiles, we weight the profile of each subhalo by the total subhalo mass.  For reference, in Fig.~\ref{fig:artemis_pdfs} we show the PDFs of the background CGM density and temperature near the dark subhaloes in all \texttt{ARTEMIS} systems, along with the PDF of the subhalo velocities (specifically its magnitude) with respect to their hosts.  Note that the variations in CGM temperature and density and subhalo velocity explored in the idealised simulations in Section \ref{sec:results} cover the vast majority of conditions sampled in \texttt{ARTEMIS}.

Fig.~\ref{fig:artemis} confirms that the signal we predicted based on analytic arguments and in idealised simulations is also present in self-consistent cosmological hydrodynamical simulations.  Furthermore, the level of the effect (tens of percent for subhaloes of $10^{10}$ M$_\odot$ down to a few percent for haloes with $10^{8-9}$ M$_\odot$) is consistent with our previously derived results.  The only significant difference, which we had anticipated, is that the spatial scale over which CGM is perturbed is significantly reduced in the cosmological simulations relative to the idealised simulations, reflecting the importance of the host halo's gravitational field in the full cosmological simulations.  Nevertheless, the signal is strongest in the central regions of the subhaloes.

The vertical dashed lines correspond to the mass-weighted mean tidal radius in each subhalo mass bin.  Specifically, for each dark subhalo we compute the radius enclosing 90\% of the bound dark matter mass.  We then compute the mean of this value in a subhalo mass bin.  In general, the extent of the CGM fluctuation (which is unbound to the dark subhalo) roughly corresponds to the tidal radius scale.  This is most evident for the pressure profiles, as pressure generally tends to be a smoother (less noisy) quantity than the density.  That the fluctuation size is similar in scale to the tidal radius strongly suggests that it is the potential of the host halo which limits the extent of the fluctuations in these simulations.

According to \texttt{ARTEMIS}, even haloes with masses as low as $10^8$ M$_\odot$ produce a discernible enhancement in the pressure and density of the CGM.  If fluctuations of this amplitude could be probed with observations, it could open the door to CGM studies placing some of the strongest constraints on the nature of dark matter.

\section{Discussion and conclusions}
\label{sec:discuss}

We have proposed a new method to search for the presence of dark subhaloes, by looking for the perturbations they source in the circumgalactic medium around normal galaxies.  We have used a combination of simple analytic arguments (Section \ref{sec:analytic}), high-resolution carefully-controlled idealised hydrodynamical simulations (Sections \ref{sec:swift} and \ref{sec:results}), and full cosmological hydrodynamical simulations (Section \ref{sec:artemis}) to calculate the expected signal and how it depends on important physical properties, including subhalo mass, CGM temperature, and relative velocity.

The main findings of this study can summarised as follows:
\begin{itemize}
    \item The compression of the CGM by dark subhaloes is a gentle, adiabatic process which, in relative terms, enhances the local CGM pressure, density, and temperature, in order of decreasing magnitude (see Fig.~\ref{fig:fiducial_profs}).  When the subhalo is at rest with respect to the CGM, the magnitude of the effect depends only on the ratio of the CGM temperature to the characteristic virial temperature of the subhalo (see Section \ref{sec:analytic}).  It does not depend on, for example, the absolute density of the CGM .
    \item In a more a realistic scenario, where the subhalo is moving with respect to the CGM, the relative velocity affects the overall morphology of the compressed region, but it does not significantly alter the magnitude of the effect (see Fig.~\ref{fig:vary_velocity}).
    \item Varying the temperature of the CGM (at fixed subhalo mass and relative velocity) not only impacts the magnitude of the effect, but also the morphology of the compressed region (see Fig.~\ref{fig:vary_temp}).  Specifically, higher CGM temperatures imply higher sound speeds so that the CGM is able to adjust its distribution more quickly in response to the presence of subhaloes, leading to a more spherical and centred (on the subhalo) compressed region.  Lower CGM temperatures lead to a slower response time which in turn leads to the enhanced region being displaced downstream from the subhalo and a more conical morphology results.
    \item Decreasing the subhalo mass/virial temperature (at fixed CGM temperature and velocity) results in a strongly reduced compression of the CGM.  Typically, a low-mass $10^8$ M$_\odot$ subhalo can enhance the CGM pressure by up to a few percent, while a high-mass $10^{10}$ M$_\odot$ subhalo can enhance it by tens of percent (see Figs.~\ref{fig:vary_msub_profs} and \ref{fig:artemis}).  As already mentioned, this depends on the CGM temperature and low-mass subhaloes would be most easily observed if embedded with a cooler phase of the CGM.  Absorption lines that pick out $10^5$ K gas (e.g., O VI) may, therefore, have the best chance for observing such systems.
    \item Not only does the presence of subhaloes compress the CGM, but it also induces local fluctuations in the velocity field, ranging from a few km/s up to $\approx$25 km/s for the range of subhalo masses and CGM temperature we have considered here.
\end{itemize}

While observations of fluctuations in the CGM of this level will be challenging, there are a number of independent methods that can be employed and cross-checked.  As the gas pressure is the most strongly affected (3-D) quantity, high-resolution and high-sensitivity tSZ effect observations, which probe the CGM pressure integrated along the line of sight, are a promising prospect.  X-ray observations, which probe the square of the gas density, should also be a promising probe.  In fact, perturbations of similar amplitude and scale to those discussed above have already been measured in deep X-ray observations of the core of the Perseus cluster (e.g., \citealt{zhuravleva2015}).  While these perturbations are likely of a different physical origin (feedback-driven turbulence), it nevertheless underscores the possibility of detecting subhalo-induced perturbations in the CGM.  The upcoming \texttt{Lynx} and \texttt{Athena} X-ray telescopes, with their exquisite angular resolution and sensitivity, should be ideal for this purpose.

Quasar absorption line studies and radio dispersion measurements linearly sample the density of the CGM and so will be less sensitive to the fluctuations in a relative sense.  A nice feature of absorption line studies, though, is that they can focus on particular phases of the CGM depending on the energy of the line.   The ability to focus on the cooler phases of the CGM in particular means gaining sensitivity to lower-mass dark subhaloes which in principle should allow for stronger constraints on the nature of dark matter.  A potential caveat is that lower temperature gas is generally thermally unstable and cools quite quickly.  On the other hand, the ubiquitous detection of, for example, O VI around normal galaxies (e.g., \citealt{tumlinson2011}) suggests a steady supply of cool, enriched gas (e.g., accretion from the intergalactic medium, or feedback-driven outflows) which can be used to probe low-mass subhaloes, as advocated above.

Lastly, while detection of individual fluctuations may be difficult in current observations, we propose that statistical measurements of the CGM fluctuation power spectra (pressure, density, and temperature) derived from wide-field observations may be a powerful alternative metric and one which can be directly compared the subhalo (or matter) power spectrum for different dark matter models.  Note, however, that fluctuations in the CGM will not only be sourced by the presence of dark subhaloes but also by a variety of other physical processes, including feedback-driven outflows\footnote{Observations could focus on particularly quiescent systems, which lack significant ongoing star formation, AGN activity, and mergers in order to minimize the non-dark subhalo signal.}.  Therefore, detailed hydrodynamical simulations will be required to interpret and isolate the cosmological signal from the astrophysical signal.  We plan to explore this question in future work.

\section*{Acknowledgements}

The authors thank the anonymous referee for helpful comments that improved the paper.  They thank Matthieu Schaller for feedback on the draft.
The authors also thank the SWIFT team for making their simulation code publicly available and easy to use.
This project has received funding from the European Research Council (ERC) under the European Union's Horizon 2020 research and innovation programme (grant agreement No 769130).  The research in this paper made use of the SWIFT open-source simulation code (\url{http://www.swiftsim.com}, \citealt{schaller2018}) version 0.8.5.  This work used the DiRAC@Durham facility managed by the Institute for Computational Cosmology on behalf of the STFC DiRAC HPC Facility. The equipment was funded by BEIS capital funding via STFC capital grants ST/P002293/1, ST/R002371/1 and ST/S002502/1, Durham University and STFC operations grant ST/R000832/1. DiRAC is part of the National e-Infrastructure.

\section*{Data availability}
The data underlying this article may be shared on reasonable request to the corresponding author.




\bibliographystyle{mnras}
\bibliography{refs} 



\appendix

\section{Hydro solver and resolution}
\label{sec:hydro_study}

Here we demonstrate the robustness of our results and conclusions to the choice of scheme used to solve the hydrodynamic equations, as well as to the adopted mass resolution of the simulations.

\begin{figure}
    \includegraphics[width=\columnwidth]{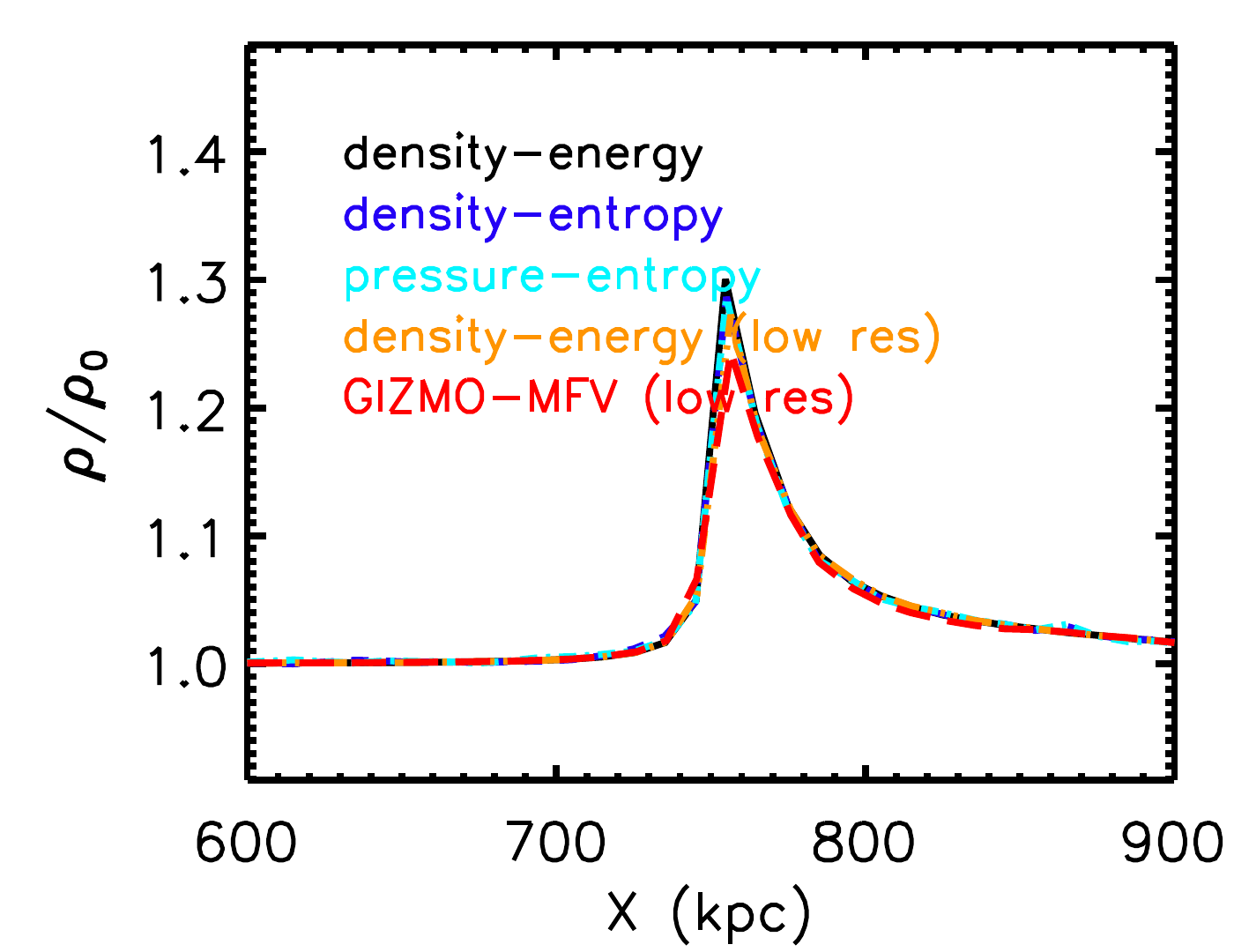}
    \caption{1-D profiles of the CGM density along the $x$-axis for the fiducial setup as the hydro solver and resolution of the idealised simulations are varied.}
    \label{fig:hydro}
\end{figure}

In Fig.~\ref{fig:hydro} we show 1-D profiles of the CGM density along the $x$-axis for the fiducial setup as the hydro solver and resolution of the idealised simulations are varied.  At fixed resolution, the default density--energy scheme (SPHENIX) yields virtually identical results to the Gadget-2 density--entropy scheme and the pressure--entropy scheme of \citet{hopkins2013}.  Reducing the mass resolution by a factor of ten for the SPHENIX case, such that there are a factor of 10 fewer particles in the box, we also see largely consistent results with the fiducial simulation, although the lower resolution simulation yields a slightly lower peak density due to enhanced smoothing.  At this lower resolution, we also experimented with the GIZMO-like `meshless finite volume' (MFV) scheme \citep{hopkins2015}, which is significantly more computationally expensive than the other schemes, at least within SWIFT.  We find largely consistent results with the default SPHENIX scheme at the same resolution.

\section{Varying the CGM density in the idealised simulations}
\label{sec:dens_dependence}

The analytic calculations in Section \ref{sec:analytic} suggest that there ought to be no explicit dependence of the (relative) enhancement of the CGM thermodynamic quantities on the background CGM density.  Here we use the idealised simulations described in Section \ref{sec:results} to explicitly test this idea.

\begin{figure}
    \includegraphics[width=\columnwidth]{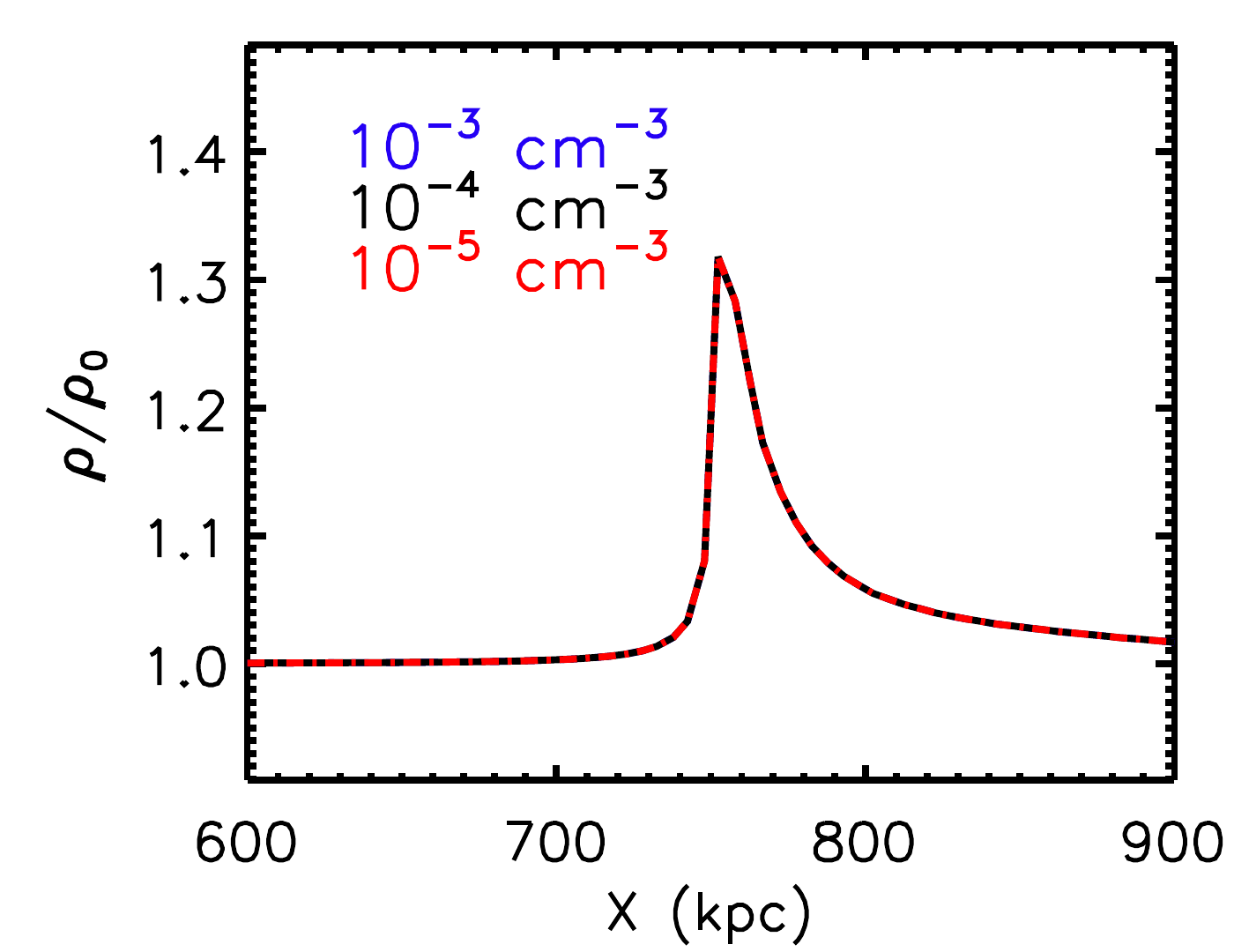}
    \caption{1-D profiles of the CGM density along the $x$-axis for the fiducial setup as the background density of the CGM is varied.}
    \label{fig:vary_dens}
\end{figure}

In Fig.~\ref{fig:vary_dens} we show 1-D profiles of the CGM density along the $x$-axis for the fiducial setup as the background density of the CGM is varied by an order of magnitude in both directions from the fiducial value of $10^{-4}$ cm$^{-3}$.  All three curves lie precisely on top of each other, indicating there is a negligible dependence on the background density in the idealised simulations, in agreement with the analytic analysis.

An important caveat to bear in mind, though, is that the idealised simulations and the analytic calculations do not account for the effects of radiative cooling.  As discussed in Section \ref{sec:results}, if pre-perturbed CGM already has a high density and, therefore, a relatively short cooling time, a further enhancement introduced by an approaching dark subhalo could potentially result in run-away cooling.  It is interesting to think that such a mechanism could be at least partly responsible for the origin of ``high-velocity clouds'' observed around the Milky Way.  Detailed analysis of simulations such as \texttt{ARTEMIS} would be helpful for exploring such a scenario, but we leave this for future work.  

\bsp	
\label{lastpage}
\end{document}